\documentclass[a4paper,12pt]{article}

\usepackage{jheppub}
\usepackage{slashed}
\usepackage{subfig}
\usepackage{xcolor}
\usepackage{booktabs}
\usepackage{adjustbox}
\usepackage{dsfont}
\usepackage{ulem}
\usepackage{footmisc}

\makeatletter
\gdef\@fpheader{}
\makeatother

\newcommand{\SM}{{\rm SM}}
\newcommand{\NP}{{\rm NP}}
\newcommand{\MFV}{{\rm MFV}}
\newcommand{\DM}{{\rm DM}}

\newcommand{\MeV}{{\,\rm MeV}}
\newcommand{\GeV}{{\,\rm GeV}}
\newcommand{\TeV}{{\,\rm TeV}}

\newcommand{\fb}{{\, \rm fb}}
\newcommand{\ab}{{\, \rm ab}}

\newcommand{\mB}{{\mathcal B}}
\newcommand{\mO}{{\mathcal O}}
\newcommand{\mC}{{\mathcal C}}
\newcommand{\mQ}{{\mathcal Q}}
\newcommand{\mL}{{\mathcal L}}

\newcommand{\sfA}{{\mathsf A}}
\newcommand{\sfB}{{\mathsf B}}

\newcommand{\hc}{\mathrm{h.c.}}

\newcommand{\inv}{\text{inv}}

\allowdisplaybreaks[4]

\title{\Large Deciphering the Belle II data on $\boldsymbol{B\to K \nu \bar\nu}$ decay in the (dark) SMEFT with minimal flavour violation}

\author[a]{Biao-Feng Hou,}
\emailAdd{resonhou@zknu.edu.cn}

\author[a,b]{Xin-Qiang Li,}
\emailAdd{xqli@ccnu.edu.cn}

\author[a]{Meng Shen,}
\emailAdd{shenmeng@mails.ccnu.edu.cn}

\author[a,c]{Ya-Dong Yang}
\emailAdd{yangyd@ccnu.edu.cn}

\author[a]{and Xing-Bo Yuan}
\emailAdd{y@ccnu.edu.cn}

\affiliation[a]{Institute of Particle Physics and Key Laboratory of Quark and Lepton Physics~(MOE),\\
Central China Normal University, Wuhan, Hubei 430079, China}
\affiliation[b]{Center for High Energy Physics, Peking University, Beijing 100871, China}
\affiliation[c]{Institute of Particle and Nuclear Physics, Henan Normal University, Xinxiang 453007, China}

\abstract{{\small Recently, the Belle II collaboration announced the first measurement of the branching ratio $\mathcal B(B^+\to K^+\nu\bar\nu)$, which is found to be about $2.7\sigma$ higher than the Standard Model~(SM) prediction. We decipher the data with two new physics scenarios: the underlying quark-level $b\to s \nu\bar\nu$ transition is, besides the SM contribution, further affected by heavy new mediators that are much heavier than the electroweak scale, or amended by an additional decay channel with undetected light final states like dark matter or axion-like particles. These two scenarios can be most conveniently analyzed in the SM effective field theory (SMEFT) and the dark SMEFT (DSMEFT) framework, respectively. We consider the flavour structures of the resulting effective operators to be either generic or satisfy the minimal flavour violation (MFV) hypothesis, both for the quark and lepton sectors. In the first scenario, once the MFV assumption is made, only one SM-like low-energy effective operator induced by the SMEFT dimension-six operators can account for the Belle II excess, whose parameter space is, however, excluded by the Belle upper bound of the branching ratio $\mathcal B(B^0\to K^{* 0}\nu\bar\nu)$. In the second scenario, it is found that the Belle II excess can be accommodated by 22 of the DSMEFT operators involving one or two scalar, fermionic, or vector dark matters as well as axion-like particles. These operators also receive dominant constraints from the $B^0\to K^{*0}+\rm inv$ and $B_s\to\rm inv$ decays. Once the MFV hypothesis is assumed, the number of viable operators is reduced to 14, and the $B^+\to\pi^+ +\rm inv$ and $K^+\to\pi^+ +\rm inv$ decays start to put further constraints on them. Within the parameter space allowed by all the current experimental data, the $q^2$ distributions of the $B\to K^{(*)}+\rm inv$ decays are then studied for each viable operator. We find that the resulting prediction of the operator $\mathcal Q_{q\chi}=\big(\bar q_p\gamma_\mu q_r\big)\big(\bar\chi\gamma^\mu\chi\big)$ with a fermionic dark matter mass $m_\chi\approx 700\,\rm MeV$ can closely match the Belle II event distribution in the bins $2\leq q^2\leq 7\,{\rm GeV}^2$. In addition, we, for the first time, calculate systematically the longitudinal polarization fraction $F_L$ of $K^\ast$ in the $B \to K^* + \inv$ decays within the DLEFT. By combining the decay spectra and $F_L$, almost all the DSMEFT operators are found to be distinguishable from each other. Finally, the future prospects at Belle II, CEPC and FCC-ee are also discussed for some of these FCNC processes.}}

\begin{document}

\date{}
\maketitle

\section{Introduction}

The flavour-changing neutral current (FCNC) processes are highly suppressed by the Glashow-Iliopoulos-Maiani (GIM) mechanism in the Standard Model (SM)~\cite{Glashow:1970gm}, making them sensitive probes of new physics (NP) beyond the SM. In particular, the $B \to K^{(\ast)} \nu \bar\nu$ decays are one of the cleanest channels within the SM and most suitable for indirect NP searches~\cite{Altmannshofer:2009ma,Buras:2014fpa, Becirevic:2023aov} (see also refs.~\cite{Felkl:2021uxi,Bause:2021cna,He:2021yoz,Browder:2021hbl,Chen:2023mep,Chen:2023wpb} for recent studies on these as well as other related decays in the context of NP, but before the recent Belle II announcement~\cite{Belle-II:2023esi}).

Recently, using the dataset with an integrated luminosity of $362\fb^{-1}$, the Belle II collaboration reported the first evidence of the $B^+ \to K^+ + \nu \bar\nu$ decay with a branching ratio~\cite{Belle-II:2023esi}
\begin{align}\label{eq:exp:Belle II}
    \mB(B^+ \to K^+ \nu \bar \nu)_{\text{Belle II}} & = \left[23 \pm 5(\text{stat})^{+5}_{-4}(\text{syst}) \right] \times 10^{-6},
\end{align}
which is obtained by combining the inclusive and hadronic tagging results. Interestingly, this measurement shows about $2.7\sigma$ higher than the SM prediction $(4.16\pm 0.57)\times 10^{-6}$, which is based on the calculations in refs.~\cite{Altmannshofer:2009ma,Buras:2014fpa,Becirevic:2023aov} but with updated input parameters. In the future, the Belle II with just $5 \ab^{-1}$ integrated luminosity is expected to measure this decay with $20 \sim 30\%$ uncertainty relative to the SM value~\cite{Belle-II:2018jsg,Belle-II:2022cgf}, and could therefore confirm or exclude such an excess. Nevertheless, it has already motivated numerous studies of possible NP explanations~\cite{Athron:2023hmz,Bause:2023mfe,Allwicher:2023xba,Felkl:2023ayn,Abdughani:2023dlr,Dreiner:2023cms,He:2023bnk,Berezhnoy:2023rxx,Datta:2023iln,Altmannshofer:2023hkn,McKeen:2023uzo,Fridell:2023ssf,Ho:2024cwk,Chen:2024jlj,Gabrielli:2024wys,Li:2024thq,Chen:2024cll}.

As an interesting feature, the Belle II result, once the $2\sigma$ range in eq.~(\ref{eq:exp:Belle II}) is considered, would be higher than the SM prediction by a factor of $3\sim 8$. Such a large excess motivates us to ask the following questions: why has such a large NP effect not shown up or will it be found in other $b \to s$ processes, such as the $B \to K^\ast \nu \bar\nu$ and $B_s \to \nu \bar\nu$ decays~\cite{Workman:2022ynf,Alonso-Alvarez:2023mgc}? In this paper, we aim to answer these questions by performing a detailed model-independent analysis of possible NP effects on these and related FCNC processes in the Effective Field Theory (EFT) framework.

The $B^+ \to K^+ \nu \bar\nu$ excess reported by Belle II could be explained by introducing heavy NP particles that are much heavier than the electroweak scale and hence contribute to the $b \to s \nu \bar\nu$ transition as virtual mediators. In this case, the Standard Model Effective Field Theory (SMEFT) provides a well-suited model-independent framework to study such kinds of NP effects~\cite{Brivio:2017vri,Isidori:2023pyp}. Here, all the NP effects are encoded in the Wilson coefficients of the higher-dimensional operators that are invariant under the SM gauge group. Below the electroweak scale, the heavy SM particles like the massive bosons and the top quark also decouple, and the dynamics is then described by the Low-energy Effective Field Theory (LEFT)~\cite{Buchalla:1995vs,Jenkins:2017jig}, which consists of, besides the QCD and QED Lagrangians for light SM fermions, a set of higher-dimensional operators compatible with the QCD and QED gauge symmetries. The Wilson coefficients of these higher-dimensional operators encode all the physics related to the heavy SM and NP degrees of freedom. Starting with the SMEFT, one can also fix these low-energy Wilson coefficients by perform a matching between these two EFTs at the electroweak scale~\cite{Jenkins:2017jig,Dekens:2019ept,Aebischer:2015fzz}. Such a procedure has been extensively applied to various $B$-meson and kaon decays~\cite{Buchalla:1995vs,Jenkins:2017jig}. In this paper, we will interpret the Belle II excess in terms of the SMEFT operators and consider the LEFT as the low-energy limit of SMEFT to calculate the relevant $B$-meson and kaon decays.

The Belle II excess could also be accommodated by amending the $B^+ \to K^+ \nu \bar\nu$ decay by an additional decay channel with undetected light final states like dark matter (DM) or axion-like particles (ALP). To mimic the signal, these new light states must couple to the SM quark sector and be sufficiently long-lived to escape the detector. The potential candidates are new light particles that are singlet under the SM gauge group. In order to study the interactions of such new states with the SM quark sector model-independently, it is also convenient to work in an EFT framework, such as the DM EFT~\cite{DelNobile:2011uf,Baumgart:2022vwr} without any stabilizing symmetry on the DM fields.\footnote{In this work, we do not impose any symmetry to stabilize the DM fields in the EFT framework. Thus, the DM fields we are considering are actually just new singlets under the SM gauge group.} In this respect, the SMEFT can be extended by including the scalar, fermionic or vector DM fields, resulting in the so-called Dark SMEFT (DSMEFT)~\cite{Duch:2014xda,Kamenik:2011vy,Brod:2017bsw,Criado:2021trs,Aebischer:2022wnl,Song:2023jqm}. In the DSMEFT, the new operators involving the DM fields are invariant under the SM $SU(3)_C \otimes SU(2)_L \otimes U(1)_Y$ gauge group and the electroweak symmetry is broken by the usual Higgs mechanism. The general non-redundant operator basis can be found in ref.~\cite{Aebischer:2022wnl} up to dim$=6$, and in ref.~\cite{Song:2023jqm} up to dim$=8$. In order to describe the low-energy processes below the electroweak scale, such as $d_i \to d_j + \DM$ decays, a general framework is the generalization of the LEFT to include the scalar, fermionic and vector DM fields, referred to as the Dark LEFT (DLEFT)~\cite{Aebischer:2022wnl}. In the DLEFT, the DM particles should be lighter than the electroweak scale and the effective operators are invariant under the $SU(3)_C\otimes U(1)_{\rm em}$ gauge group. A complete basis of the DLEFT operators can be found in refs.~\cite{Aebischer:2022wnl,He:2022ljo} for dim $\leq 6$, and in ref.~\cite{Liang:2023yta} for dim $\leq 7$. In this work, starting from the DSMEFT and considering the DLEFT as the low-energy limit of DSMEFT, we investigate the parameter space of each DSMEFT operator required to explain the Belle II excess, together under the constraints from the other $b \to s + \inv$ decays. The future prospects at Belle II, CEPC and FCC-ee will also be discussed for some of these FCNC processes. In addition, for completeness, our analyses also include the ALP case~\cite{Peccei:1977hh,Peccei:1977ur,Weinberg:1977ma,Wilczek:1977pj} in the EFT framework~\cite{Georgi:1986df,Brivio:2017ije,Chala:2020wvs,Bauer:2020jbp,Galda:2021hbr,Song:2023lxf}.

The above two NP scenarios proposed to account for the Belle II excess could also affect the quark-level $b \to d$ and $s \to d$ processes. In order to correlate the $b \to s$ to the $b \to d$ and $s \to d$ transitions, we have to specify the underlying flavour structures of the EFT considered~\cite{Isidori:2010kg,Altmannshofer:2022aml}. To this end, we assume the minimal flavour violation (MFV) hypothesis~\cite{Chivukula:1987py,Buras:2000dm,DAmbrosio:2002vsn} and implement it into the SMEFT and DSMEFT frameworks. According to the hypothesis, all the flavour violating currents are controlled by the SM Yukawa couplings, so that all the FCNC interactions in the quark sector are naturally suppressed by the CKM factors. Therefore, any potential large FCNCs in the SMEFT and DSMEFT can be avoided. Considering the current experimental data on the $b \to s + \inv$, $b \to d + \inv$ and $s \to d + \inv$ processes, such as the $B \to \pi + \inv$ and $K \to \pi + \inv$ decays, we derive the constraints on each operator in the SMEFT and DSMEFT with the MFV hypothesis. Implications of the resulting parameter space of each viable operator for the differential distributions of these decays are also investigated in detail, which can be tested at the Belle II with more statistics. In addition, in order to distinguish the various DSMEFT operators, we, for the first time, calculate systematically the polarization fraction $F_L$ of $K^*$ in the $B \to K^* + \inv$ decays within the DELFT.

The remainder of this paper is organized as follows. In section~\ref{sec:MFV}, we introduce the MFV hypothesis. In sections~\ref{sec:theory:nunu} and \ref{sec:theory:DM}, we investigate the various $d_i \to d_j \nu \bar\nu$ and $d_i \to d_j + \DM$ decays in the SMEFT and DSMEFT, respectively. In these EFT frameworks, the flavour structures of the effective operators are taken to be either of the most general form or satisfy the MFV hypothesis. We present our detailed numerical analyses and discussions in section~\ref{sec:numerical analysis}. Our conclusion is finally made in section~\ref{sec:conclusion}. The hadronic matrix elements involved throughout this work, such as the transition form factors and decay constants, are summarized in appendix~\ref{eq:hadronic matrix element}.

\section{Minimal Flavour Violation}
\label{sec:MFV}

In this section, we discuss the idea of MFV hypothesis and its application in the SMEFT and DSMEFT.

\subsection{Quark sector}

In the interaction eigenbasis, the SM Yukawa interactions in the quark sector are described by the Lagrangian
\begin{align}
  -\mL_Y = \bar q \, Y_d H d + \bar q \, Y_u \tilde H u + \hc,
\end{align}
where $\tilde H \equiv i \sigma_2 H^*$ with $H$ being the SM Higgs doublet. The fields $q$ and $u,d$ denote the left-handed quark doublet and the right-handed quark singlets in the SM, respectively. The Yukawa coupling matrices $Y_{u,d}$ are $3 \times 3 $ complex matrices in flavour space. In the SM, these Yukawa interactions violate the global flavour symmetry~\cite{Gerard:1982mm,Chivukula:1987py}
\begin{align}
  G_{\rm QF} = SU(3)_q \otimes SU(3)_u \otimes SU(3)_d.
\end{align}
We can formally recover this flavour symmetry by promoting the Yukawa matrices $Y_{u,d}$ to spurion fields~\cite{DAmbrosio:2002vsn}. The Yukawa spurions should transform under $G_{\rm QF}$ as
\begin{align}
  Y_u \sim \big( \boldsymbol{3}, \boldsymbol{\bar 3}, \boldsymbol{1} \big),
  \qquad
  Y_d \sim \big( \boldsymbol{3}, \boldsymbol{1}, \boldsymbol{\bar 3}\big).
\end{align}
Then, we can construct two useful basic building blocks $\sfA \equiv Y_u Y_u^\dagger$ and $\sfB\equiv Y_d Y_d^\dagger$, which transform as $(\boldsymbol{1} \oplus \boldsymbol{8}, \boldsymbol{1}, \boldsymbol{1})$ under the flavour group $G_{\rm QF}$. As will be seen later, these building blocks are useful to parameterize the quark flavour transitions in effective theories.

In the SMEFT and DSMEFT, there are four types of quark currents relevant for the down-type FCNC transitions, which read
\begin{align}\label{eq:quark current}
 \bar q  \gamma^\mu \mC q, \qquad \bar d \gamma^\mu \mC d, \qquad   \bar q \mC d, \qquad \bar q \sigma^{\mu\nu} \mC d,
\end{align}
where $\mC$ denotes the Wilson coefficient of the operator $\mQ$ and is a $3 \times 3$ matrix in quark flavour space. Here and in the following, we suppress the operator index $i$, i.e., $\mC$ denoting $\mC_i$. In the MFV hypothesis, it is assumed that at low energy the Yukawa couplings are the only source of flavour violation in the SM and beyond~\cite{DAmbrosio:2002vsn}. Technically, this hypothesis implies that the quark currents in eq.~(\ref{eq:quark current}) should be invariant under the flavour group $G_{\rm QF}$. Therefore, the Wilson coefficient $\mC$ can be written in the following form:
\begin{align}
  \mC^{\rm MFV}=
  \begin{cases}
    f(\sfA, \sfB) &     \text{for \;} \bar q  \gamma^\mu \mC q, 
    \\
    f(\sfA, \sfB) Y_d & \text{for \;} \bar q \mC d,\, \bar q \sigma^{\mu\nu} \mC d,
    \\
    \epsilon_0 \mathds{1} + Y_d^\dagger g(\sfA, \sfB) Y_d &           \text{for \;} \bar d \gamma^\mu \mC d,
  \end{cases}
\end{align}
where $\epsilon_0$ is a real constant. The functions $f(\sfA, \sfB)$ and $g(\sfA,\sfB)$ are infinite series of the building blocks $\sfA$ and $\sfB$, e.g., explicitly $f(\sfA, \sfB)\equiv \xi_{ijk \dotsm} \sfA^i \sfB^j \sfA^k \dotsb$ with real coefficients $\xi_{ijk\dotsm}$ to ensure no new sources of CP violation beyond the Yukawa couplings. By application of the Cayley-Hamilton identity, the infinite series in $f(\sfA, \sfB)$ and $g(\sfA,\sfB)$ can be resummed into a finite number of terms. Taking $f(\sfA, \sfB)$ as an example, its explicit summed expression can be written as~\cite{Colangelo:2008qp,Mercolli:2009ns,Grinstein:2023njq}
\begin{align*}
  f(\sfA,\sfB)=\epsilon_0 \mathds{1} +&\epsilon_1 \sfA  + \epsilon_3 \sfA^2 + \epsilon_5 \sfA\sfB  + \epsilon_7\sfA\sfB\sfA  +  \epsilon_{10}\sfA \sfB^2 + \epsilon_{12}\sfA^2\sfB^2 +\epsilon_{14}\sfB^2\sfA\sfB+\epsilon_{15}\sfA\sfB^2\sfA^2 \nonumber\\
  +& \epsilon_2 \sfB+ \epsilon_4 \sfB^2+ \epsilon_6 \sfB\sfA +\epsilon_{9} \sfB\sfA\sfB +\epsilon_8 \sfB\sfA^2 + \epsilon_{13}\sfB^2\sfA^2 + \epsilon_{11}\sfA\sfB\sfA^2   +\epsilon_{16}\sfB^2\sfA^2\sfB.
\end{align*}
During the resummation, higher-order polynomials of $\sfA$ and $\sfB$ contribute to the coefficients $\epsilon_i$, and make them complex generally. However, it is found that the imaginary parts of the coefficients are tiny (e.g., $\text{Im}\epsilon_i \propto |\text{Tr}(\sfA^2 \sfB \sfA \sfB^2)|\ll 1$) and can be therefore neglected in the numerical analysis~\cite{Colangelo:2008qp,Mercolli:2009ns,He:2014uya,He:2014fva,He:2014efa}. In the following, we assume the coefficients $\epsilon_i$ are of the same orders of magnitude. Therefore, only the terms built by $\sfA$ are kept, while all the terms involving $\sfB$ are neglected, as the spurion $\sfB$ is suppressed by $\mO(\hat{\lambda}_d^2)$, where $\hat{\lambda}_d$ denotes the diagonal Yukawa coupling matrix for down-type quarks defined by eq.~\eqref{eq:lambdaudhat}. Finally, we obtain the approximation~\cite{Chiang:2017hlj}
\begin{align}
  f(\sfA, \sfB) \approx \epsilon_0 \mathds{1} + \epsilon_1 \sfA + \epsilon_2 \sfA^2,
\end{align}
where the coefficients $\epsilon_{0,1,2}$ are free real parameters. Similarly, being suppressed by $\mO(\hat{\lambda}_d^2$), the approximation $Y_d^\dagger g(\sfA,\sfB)Y_d \approx 0$ can be taken.

In the MFV framework discussed above, the quark currents are all given in the interaction eigenbasis. In order to rotate to the fermion mass eigenbasis, the following transforms should be performed:
\begin{align}
  u_L \to U_u u_L, \qquad  u_R \to W_u u_R, \qquad 
  d_L \to U_d d_L, \qquad d_R \to W_d d_R,
\end{align}
where $U_{u,d}$ and $W_{u,d}$ are $3\times 3$ unitary matrices introduced to diagonalize the Yukawa matrices
\begin{align} \label{eq:lambdaudhat}
  U_d^\dagger Y_d W_d = \hat{\lambda}_d, \qquad U_u^\dagger Y_u W_u =\hat{\lambda}_u.
\end{align}
Therefore, the CKM matrix is given by $V=U_u^\dagger U_d$. In the following, we turn to the fermion mass eigenbasis and derive explicit expressions of the MFV couplings.

The current $\bar q \gamma^\mu \mC q$ involves the following down-type quark interactions in the mass eigenbasis:
\begin{align}
\bar d_L U_d^\dagger \gamma_\mu \mC^\MFV U_d d_L \approx   \bar d_L \gamma^\mu V^\dagger \big(\epsilon_0 + \epsilon_1 \hat{\lambda}_u^2 + \epsilon_2 \hat{\lambda}_u^4 \big) V d_L,
\end{align}
where $d_L$ denotes now the left-handed down-type quark in the mass eigenbasis, and $\hat{\lambda}_u$ denotes the diagonal Yukawa coupling matrix for up-type quarks, whose diagonal elements are given by $\lambda_{u_i} = \sqrt{2} m_{u_i} /v$ with the vacuum expectation value $v = 246\GeV$. It is noted that, due to the large hierarchy among the diagonal elements $\lambda_{u_i}$, the matrices $\epsilon_1 \hat{\lambda}_u^2$ and $\epsilon_2 \hat{\lambda}_u^4$ have almost the same structures. Therefore, one can obtain the following approximation~\cite{Zhang:2018nmy}:
\begin{align}\label{eq:MFV:approx}
  V^\dagger (\epsilon_1 \hat{\lambda}_u^2 + \epsilon_2 \hat{\lambda}_u^4) V \approx V^\dagger (\epsilon_1 \hat{\lambda}_u^2) V\equiv \epsilon_1 \Delta_q,
\end{align}
which is equivalent to taking the approximation $\lambda_t^4- \lambda_{u,c}^4 \approx \lambda_t^2 (\lambda_t^2-\lambda_{u,c}^2)$ and redefining $\epsilon_1 + \epsilon_2 \lambda_t^2 \to \epsilon_1$.\footnote{Using the unitarity of the CKM matrix, one can obtain $
  [V^\dagger (\epsilon_1 \hat{\lambda}_u^2 + \epsilon_2 \hat{\lambda}_u^4) V]_{ij}
  = \lambda_t^2(\epsilon_1 + \epsilon_2 \lambda_t^2) \delta_{ij} - (\lambda_t^2-\lambda_u^2)[\epsilon_1 + \epsilon_2 (\lambda_t^2 + \lambda_u^2)]V_{ui}^*V_{uj} -  (\lambda_t^2-\lambda_c^2)[\epsilon_1 + \epsilon_2 (\lambda_t^2 + \lambda_c^2)]V_{ci}^*V_{cj}
  \approx \lambda_t^2(\epsilon_1 + \epsilon_2 \lambda_t^2) \delta_{ij} - (\lambda_t^2-\lambda_u^2)[\epsilon_1 + \epsilon_2 \lambda_t^2]V_{ui}^*V_{uj} -  (\lambda_t^2-\lambda_c^2)[\epsilon_1 + \epsilon_2 \lambda_t^2 ]V_{ci}^*V_{cj}
  = (\epsilon_1 + \epsilon_2 \lambda_t^2)[V^\dagger \hat{\lambda}_u^2 V]_{ij}$, where the approximation $\lambda_t^2+\lambda_{u,c}^2 \approx \lambda_t^2$ have been used.} For the observables investigated in this paper, we have verified that the numerical differences caused by this approximation are negligible. Finally, by defining the basic FCNC couplings
\begin{align}
  \Delta_q = V^\dagger \hat{\lambda}_u^2 V,
\end{align}
we arrive at
\begin{align}\label{eq:MFV:vector current:dL}
  \bar d_L U_d^\dagger \gamma_\mu \mC^\MFV U_d d_L \approx   \bar d_L \gamma^\mu \big(\epsilon_0 \mathds{1} + \epsilon_1 \Delta_q) d_L.
\end{align}

Similarly, from the current $\bar q \,\mC d$, the following scalar interactions among the down-type quarks are obtained in the mass eigenbasis:
\begin{align}\label{eq:MFV:scalar current}
  \bar d_L U_d^\dagger \mC^\MFV W_d d_R
  \approx \bar d_L V^\dagger \big( \epsilon_0 \mathds{1} + \epsilon_1 \hat{\lambda}_u^2 + \epsilon_2 \hat{\lambda}_u^4 \big) V \hat{\lambda}_d d_R
  \approx \bar d_L \big( \epsilon_0 \hat{\lambda}_d + \epsilon_1 \Delta_q \hat{\lambda}_d \big) d_R,
\end{align}
where $d_R$ is the right-handed down-type quark in the mass eigenbasis, and the diagonal elements of the diagonal Yukawa coupling matrix for down-type quarks are given by $\lambda_d^i = \sqrt{2}m_{d_i} /v$. Here the approximation in eq.~(\ref{eq:MFV:approx}) has been used. In addition, due to the same flavour structure as of the scalar interactions, the tensor interactions $\bar d_L U_d^\dagger \sigma^{\mu\nu} \mC^{\rm MFV} W_d d_R$ resulting from the current $\bar q \sigma^{\mu\nu} \mC^{\rm MFV} d$ have the same MFV coupling as in eq.~\eqref{eq:MFV:scalar current}.

For the vector interactions originating from the current $\bar d \gamma^\mu \mC d$, the MFV coupling in the mass eigenbasis is given by
\begin{align}\label{eq:MFV:vector current:dR}
 \bar d_R W_d^\dagger \gamma^\mu \mC^\MFV W_d d_R \approx \bar d_R \gamma^\mu ( \epsilon_0 \mathds{1}) d_R.
\end{align}
This implies that the vector interactions among the right-handed down-type quarks are approximately flavour diagonal and universal. Thus, the FCNC vector interactions among the right-handed down-type quarks are forbidden in the MFV scenario.

As a summary, the MFV couplings for various down-type quark currents are given in the mass eigenbasis by
\begin{align}
  \mC_i^\MFV =
  \begin{cases}
    \epsilon_0^i \mathds{1} + \epsilon_1^i \Delta_q
    & \text{for }  \bar d_L  \gamma^\mu \mC_i d_L, 
    \\
    \epsilon_0^i \hat{\lambda}_d + \epsilon_1^i \Delta_q \hat{\lambda}_d
    & \text{for }  \bar d_L \mC_i d_R,\, \bar d_L \sigma^{\mu\nu} \mC_i d_R,
    \\
    \epsilon_0^i \mathds{1}
    & \text{for }  \bar d_R \gamma^\mu \mC_i d_R.
  \end{cases}
\end{align}
It is then clear that only two types of MFV couplings, $\Delta_q $ and $\Delta_q \hat{\lambda}_d$, can generate FCNC interactions among the down-type quarks and their numerical values are given, respectively, by
\begin{align}\label{eq:MFV:num:1}
  \Delta_q 
  = 
  \begin{pmatrix}
     0.8       &   -3.3-1.5i &    79.3+35.4i  \\
    -3.3+ 1.5i &   16.6      &  -397.5+ 8.1i \\
    79.3-35.4i & -397.5-8.1i &  9839.0
  \end{pmatrix}
  \times 10^{-4},
\end{align}
and
\begin{align}\label{eq:MFV:num:2}
  \Delta_q \hat{\lambda}_d
  = 
  \begin{pmatrix}
     0.0021         &  -0.18-0.08i     & 191.3+85.4i \\
    -0.009+0.004i &   0.88         & -958.7+19.6i \\
     0.21-0.10i & -21.1-0.4i & 23728.1 \\
   \end{pmatrix}
  \times 10^{-6}.
\end{align}
One can see that a large hierarchy exists among the down-type FCNC interactions, i.e., $\mC_{bs}^\MFV \gg \mC_{bd}^\MFV \gg \mC_{sd}^\MFV$. It is also seen that the flavour-conserving couplings $\mC_{bb}^\MFV$ is more than one order of magnitude larger than all the other couplings.

\subsection{Lepton sector}

One can also apply the MFV hypothesis to the lepton sector. However, since the mechanism of neutrino mass generation is still unknown, there are different approaches to formulate the leptonic MFV~\cite{Cirigliano:2005ck,Cirigliano:2006su,Davidson:2006bd,Branco:2006hz,Gavela:2009cd,Alonso:2011jd,AristizabalSierra:2012myh}. Here, we consider the realization of leptonic MFV within the so-called minimal field content~\cite{Cirigliano:2005ck,Cirigliano:2006su}, in which the neutrino masses are generated by the Weinberg operator. In this case, the Yukawa interactions in the lepton sector can be written as
\begin{align}
 - \Delta \mL = \bar e\, Y_e H^\dagger l + \frac{1}{2 \Lambda_{\rm LN}} \big( \bar l ^c \tau_2 H \big) Y_\nu \big( H^T \tau_2 l \big) + \hc,
\end{align}
where $l$ denotes the left-handed lepton doublet with the charge conjugated field given by $l^c =-i \gamma_2 l^\ast$, and $e$ is the right-handed charged lepton singlet. $\Lambda_{\rm LN}$ denotes the breaking scale of the lepton number symmetry $U(1)_{\rm LN}$. $Y_e$ and $Y_\nu$ stand for the $3 \times 3$ Yukawa coupling matrices in flavour space. In the absence of these Yukawa couplings, the lepton sector respects the flavour symmetry
\begin{align}
  G_{\rm LF} = SU(3)_l \otimes SU(3)_e.
\end{align}
Analogous to the quark sector, this flavour symmetry can be restored by promoting the Yukawa couplings $Y_{e,\nu}$ to spurion fields. Then, the relevant building blocks are $\sfA_\ell = Y_\nu^\dagger Y_\nu$ and $\sfB_\ell = Y_e^\dagger Y_e$, which transform as $(\textbf{1}\oplus \textbf{8}, \textbf{1})$ under the lepton flavour group $G_{\rm LF}$.

For the three SMEFT operators, $\mQ_{lq}^{(1,3)}$ and $\mQ_{ld}$, relevant to the $B^+ \to K^+ \nu \bar \nu$ decay, they contain the lepton current
\begin{align}
  \bar l \gamma^\mu\mC l,
\end{align}
where $\mC$ denotes the Wilson coefficient of the operator $\mQ$ and is a $3 \times 3$ matrix in lepton flavour space. In the MFV hypothesis, analogous to the quark current, the Wilson coefficient $\mC$ should take the form $\mC_\MFV = h(\sfA_\ell, \sfB_\ell)$, where $h(\sfA_\ell, \sfB_\ell)$ is a finite polynomial of $\sfA_\ell$ and $\sfB_\ell$. After neglecting all the terms involving $\sfB_\ell$, which are suppressed by the small lepton Yukawa couplings $Y_e$, we obtain
\begin{align}\label{eq:MFV:V:l}
  \mC_\MFV \approx \kappa_0 + \kappa_1 \sfA_\ell + \kappa_2 \sfA_\ell^2,
\end{align}
where the coefficients $\kappa_{0,1,2}$ are free real parameters. In the numerical analysis, we keep only the leading lepton flavour violation term $\sfA_\ell$ for simplicity, i.e., $\kappa_2=0$. Turning to the lepton mass eigenbasis, the current $\bar l \gamma^\mu \mC l$ gives in the MFV hypothesis the following interactions:
\begin{align}\label{eq:MFV:vector current:l}
  \bar e_L \gamma^\mu(\kappa_0 \mathds{1} + \kappa_0 \Delta_\ell) e_L + \bar \nu_L \gamma^\mu(\kappa_0 \mathds{1} + \kappa_0 \hat{\lambda}_\nu^2) \nu_L,
\end{align}
where the basic LFV coupling $\Delta_\ell$ can be obtained from $\sfA_\ell$ and takes the form
\begin{align}
  \Delta_\ell = U \hat{\lambda}_\nu^2 U^\dagger,
\end{align}
where $U$ is the PMNS matrix. Here $e_L$ and $\nu_L$ denote the left-handed charged lepton and neutrino in the mass eigenbasis, respectively. $\hat{\lambda}_\nu \equiv 2m_\nu\Lambda_{\rm LN} /v^2$ stands for the diagonal effective neutrino Yukawa coupling and $m_\nu$ the diagonal neutrino mass matrix. As can be seen from eq.~\eqref{eq:MFV:vector current:l}, the lepton vector current does not induce neutral LFV in the leptonic MFV with minimal field content.\footnote{This is, however, not true for the most generic leptonic MFV. For example, in the leptonic MFV with type-I seesaw mechanism, the basic LFV coupling $\Delta_\ell = U \hat{\lambda}_\nu U^\dagger$ should be replaced by $U\hat{\lambda}_\nu^{1/2} O O^\dagger \hat{\lambda}_\nu^{1/2}U^\dagger$, where $O$ is a general complex orthogonal matrix satisfying $O O^T = \mathds{1}$~\cite{Chiang:2017hlj}. Therefor, any non-diagonal matrix $OO^\dagger$ can induce neutral LFV.}

As discussed above, the charged LFV interactions are governed by the MFV coupling $\Delta_\ell$. Numerically, we obtain
\begin{align}
  \Delta_\ell^{\rm NO}  =
  \begin{pmatrix}
               -0.19-0.01i \; &             -0.25-0.02i \; & \hphantom{-} 0.31-0.04i \\
    \hphantom{-}0.12+0.01i \; & \hphantom{-} 0.28-0.00i \; & \hphantom{-} 0.29+0.04i \\
               -0.37-0.01i \; & \hphantom{-} 0.21-0.05i \; &             -0.03+0.01i
  \end{pmatrix},
\end{align}
for normal ordering (NO), and
\begin{align}
  \Delta_\ell^{\rm IO} =
  \begin{pmatrix}
    0.21+0.09i \; &            -0.34+0.05i \; & \hphantom{-}0.03+0.11i \\ 
    0.31+0.12i \; & \hphantom{-}0.19+0.00i \; &            -0.15-0.14i \\
    0.12-0.02i \; & \hphantom{-}0.04-0.19i \; & \hphantom{-}0.34-0.10i
  \end{pmatrix}, 
\end{align}
for inverted ordering (IO) of neutrino masses. Here $\Lambda_{\rm LN}=10^{14}\GeV$ and the lightest neutrino mass $m_{\nu_1}=0.2\,{\rm eV}$ have been taken. One can see that the magnitudes of all the matrix elements of $\Delta_\ell$ are of similar size except for $(\Delta_\ell^{\rm NO})_{33}$, which has much smaller magnitude than the other couplings.

Let us finally make a comment on the DSMEFT with MFV. In general, similar to the SM fermions, the DM particles can also appear in a form of multiple generations. In this case, the SM flavour symmetry could be extended to include the global symmetry associated with the DM multiplet and the MFV hypothesis can be generalized to the DM sector~\cite{Batell:2011tc,Lopez-Honorez:2013wla,Agrawal:2014aoa,Chen:2015jkt,Acaroglu:2021qae}. In this paper, we focus on a minimal DM sector, which contains only one or two DM generations, and leave a detained study of the DM with MFV for future work.

\section{$\boldsymbol{M_1 \to M_2 + \nu \bar \nu}$}
\label{sec:theory:nunu}

In this section, we first summarize the SMEFT framework with the most general flavour structure as well as in the MFV hypothesis. Then, the NP effects on  various rare FCNC decays are discussed.

\subsection{SMEFT}
\label{sec:SMEFT}

If new particles are much heavier than the electroweak scale and the electroweak symmetry breaking is realized linearly, their effects can be parameterized by a series of higher dimensional gauge-invariant operators in the SMEFT~(see refs.~\cite{Brivio:2017vri,Isidori:2023pyp} for recent reviews). The SMEFT Lagrangian at dim-6 takes the form~\cite{Buchmuller:1985jz,Grzadkowski:2010es}
\begin{align}\label{eq:SMEFT:L}
	\mathcal{L}_{\rm SMEFT} =  \sum_i \frac{\mC_i}{\Lambda^2} \mQ_i,
\end{align}
where $\Lambda$ denotes the NP scale. The dim-6 effective operators $\mQ_i$ are built out of the SM fields and respect the SM gauge symmetry $SU(3)_C \otimes SU(2)_L \otimes U(1)_Y$, with $\mC_i$ the corresponding dimensionless Wilson coefficients.

For the $b \to s \nu \bar\nu$ transition, the relevant operators are given by
\begin{align} \label{eq:SMEFT:operators}
  \mQ_{lq}^{(1)} &= \big ( \bar l_p\gamma^\mu l_r\big ) \big (\bar q_s \gamma_\mu q_t \big )\,,
  &
  \mQ_{Hq}^{(1)} & = \big( H^\dagger i \overleftrightarrow{D}_\mu H \big) \big (\bar q_p \gamma^\mu q_r \big)\,,
  \\\nonumber
  \mQ_{lq}^{(3)} &=\big (\bar l_p\gamma^\mu \tau^ I l_r\big ) \big (\bar q_s \tau^ I\gamma_\mu q_t\big )\,,
  &
  \mQ_{Hq}^{(3)} & = \big( H^\dagger i \overleftrightarrow{D}^I_\mu H \big) \big(\bar q_p \tau^I \gamma^\mu  q_r \big),
  \\ \nonumber
  \mQ_{ld} &=\big ( \bar l_p\gamma^\mu  l_r\big ) \big ( \bar d_s\gamma_\mu d_{t}\big ),
  &
  \mQ_{Hd} & =  \big( H^\dagger i \overleftrightarrow{D}_\mu H \big) \big (\bar d_p \gamma^\mu d_r \big),
\end{align}
where $D_\mu = \partial_\mu + i g_s T^a G^a_\mu + i g_2 t^I W_\mu^I + i g_1 Y B_\mu $ denotes the covariant derivative under the SM gauge symmetry. Here $T^a = \lambda^a/2$ and $t^I = \tau^I /2$ are the $SU(3)$ and $SU(2)$ generators, where $\lambda^a$ and $\tau^I$ stand for the Gell-Mann and Pauli matrices, respectively. $l$ and $q$ denote the left-handed lepton and quark doublets respectively, while $d$ is the right-handed down-type quark singlet, all being given in the mass eigenbasis. The generation indices of the SM fields are characterized by $p$, $r$, $s$, $t$.

The operators $\mQ_{Hq}^{(1)}$, $\mQ_{Hq}^{(3)}$ and $\mQ_{Hd}$ can induce a tree-level $\bar s \gamma^\mu b Z_\mu$ coupling, and thus contribute to the $b \to s \nu \bar \nu$ transition. However, the same $\bar s \gamma^\mu b Z_\mu$ interaction also affects the $b \to s \ell^+ \ell^-$ processes~\cite{Altmannshofer:2009ma,Buras:2014fpa}. Due to the $SU(2)_L$ gauge invariance, its contributions to these two processes are of equal magnitudes and lepton flavour universal. Thus, it is impossible to explain the Belle II data on the branching ratio $\mB(B^+ \to K^+ \nu \bar\nu)$ while satisfying the stringent constraints from the precisely measured $b \to s \ell^+ \ell^-$ observables, e.g., the branching ratio $\mB(B_s \to \mu^+ \mu^-)$. Explicitly, it is found that $\mB(B^+ \to K^+ \nu \bar \nu)$ can only be enhanced by about $20\%$ after considering other relevant experimental constraints~\cite{Allwicher:2023xba}. Therefore, we will not consider these operators and focus on the remaining three four-fermion operators in the following.

In the MFV hypothesis, by using eqs.~(\ref{eq:MFV:vector current:dL}), (\ref{eq:MFV:vector current:dR}) and (\ref{eq:MFV:vector current:l}), the relevant four-fermion operators take the following forms:
\begin{align}\label{eq:MFV:SMEFT}
  \mC_{ld}\mQ_{ld}&= \Big [ \bar e_L \gamma_\mu \left(\kappa_0 \mathds{1} + \kappa_1 \Delta_\ell\right) e_L  +  \bar\nu_L \gamma_\mu \big(\kappa_0 \mathds{1} + \kappa_1 \hat{\lambda}_\nu^2\big) \nu_L  \Big ] \Big[ \bar d_R \gamma^\mu (\epsilon_0 \mathds{1})  d_R \Big ], \\[0.15cm]
  \mC_{lq}^{(1,3)}\mQ_{lq}^{(1,3)}& \supset  \Big [ \bar e_L \gamma_\mu \left(\kappa_0 \mathds{1} + \kappa_1 \Delta_\ell\right) e_L \pm \bar\nu_L \gamma_\mu \big(\kappa_0 \mathds{1} + \kappa_1 \hat{\lambda}_\nu^2\big) \nu_L   \Big ] \Big[ \bar d_L \gamma^\mu \left(\epsilon_0 \mathds{1} +\epsilon_1 \Delta_q\right)  d_L \Big ], \nonumber
\end{align}
in the mass eigenbasis. Here we have suppressed the operator indices for the MFV parameters $\epsilon_{0,1}$ and $\kappa_{0,1}$ are suppressed. It is noted that $\mQ_{ld}$ does not involve any quark FCNC interactions, and thus has no contribution to the $b \to s \nu \bar\nu$ transition.

\subsection{Observables}
\label{sec:obs:nu}

\subsubsection{$b \to s (d) \nu \bar{\nu}$ decay}
\label{sec:b2snunu}

The $B \to K^{(*)} \nu \bar{\nu}$ decays are induced by the $b\to s \nu_i \bar\nu_j$ transitions. In the most general case, the effective weak Hamiltonian contains only two operators and can be written as~\cite{Altmannshofer:2009ma,Buras:2014fpa}
\begin{align} \label{eq:Heff:bsnunu}
  \mathcal{H}_{\rm eff}^{b \to s \nu\nu} =  -\dfrac{4 G_F}{\sqrt{2}} \frac{\alpha_e}{4 \pi} V_{tb}V_{ts}^* \sum_{i,j = e,\mu,\tau} \left( C^{\nu_i\nu_j}_L \mO_L^{\nu_i \nu_j}\,+ C^{\nu_i\nu_j}_R \mO_R^{\nu_i\nu_j}\right) +\hc,
\end{align}
with the effective operators defined by
\begin{align}
  \mO_L^{\nu_i \nu_j}= \big( \bar s \gamma_\mu P_L b \big) \big( \bar{\nu}_i  \gamma^\mu P_L \nu_j \big),
  \qquad
  \mO_R^{\nu_i \nu_j}= \big( \bar s \gamma_\mu P_R b \big) \big( \bar{\nu}_i  \gamma^\mu P_L \nu_j \big).
\end{align}
In the SM, $C_R^{\nu_i \nu_j}$ is negligible and $C_L^{\nu_i \nu_j} = C_L^\SM \delta_{ij}$ with $C_L^\SM=-12.64 \pm 0.14$~\cite{Buras:2014fpa} after including the NLO QCD corrections~\cite{Buchalla:1993bv,Misiak:1999yg,Buchalla:1998ba} and the two-loop electroweak contributions~\cite{Brod:2010hi}. In the SMEFT, by comparing eqs.~\eqref{eq:SMEFT:L} and \eqref{eq:Heff:bsnunu}, one can derive that the NP contributions to these two operators are given, respectively, by
\begin{align}\label{eq:match:CL_bsnunu}
  C_{L, \NP}^{\nu_i \nu_j}  = c_\nu \big (\big{[}\mathcal{C}_{lq}^{(1)}\big{]}_{ij23}-\big{[}\mathcal{C}_{lq}^{(3)}\big{]}_{ij23}\big ),
  \qquad
  C_{R, \NP}^{\nu_i \nu_j}  = c_\nu \big[ \mathcal{C}_{ld} \big ]_{ij23} ,
\end{align}
with the normalization factor $c_\nu \equiv (2\pi /(\alpha_e V_{tb}V_{ts}^*)) \cdot (v^2/\Lambda^2)$. In the MFV framework, by using eq.~(\ref{eq:MFV:SMEFT}), we find that
\begin{align}
  C_{L,\MFV}^{\nu_i \nu_j} & = c_\nu \Delta_q^{23} \big[ \epsilon_1^{(1)} \big(\kappa_0^{(1)} + \kappa_1^{(1)} \lambda_{\nu_i}^2 \big) - \big(  (1) \to (3) \big) \big] \delta_{ij}, \\
  C_{R,\MFV}^{\nu_i \nu_j} & = 0,
  \nonumber
\end{align}
where $\epsilon_1^{(n)}$ and $\kappa_{0,1}^{(n)}$ denote the MFV parameters associated with the operator $\mQ_{lq}^{(n)}$, with $n=1,3$. Thus, one can see that the transition $b \to s \nu \bar\nu$ in the MFV hypothesis is only induced by the effective operator $\mO_L^{\nu_i\nu_i}$, which also conserves the lepton flavour. In addition, it is straightforward to obtain the operators and Wilson coefficients for the $b \to d \nu_i \bar\nu_j$ transitions by changing the corresponding flavour indices.

Starting with the effective weak Hamiltonian in eq.~(\ref{eq:Heff:bsnunu}), we can write the dineutrino invariant mass spectrum of the $B \to K^* \nu \bar{\nu}$ decay as~\cite{Altmannshofer:2009ma,Buras:2014fpa}
\begin{align}
  \frac{d \Gamma(B \to K^* \nu \bar\nu)}{d s_B}=m_B^2 \sum_{i,j=e,\mu,\tau}\Big(\left|A^{ij}_{\perp}(s_B)\right|^2+\left|A^{ij}_{\|}(s_B)\right|^2+\left|A^{ij}_0(s_B)\right|^2\Big),
\end{align}
where $s_B=q^2 / m_B^2$, and $q^2$ denotes the dineutrino invariant mass squared in the $B$-meson rest frame. The three polarization amplitudes $A_{\perp,\|,0}^{ij}$ are functions of the $B \to K^*$ transition form factors and the Wilson coefficients, and take the form
\begin{align}
  A^{ij}_\perp(s_B) &=2 \sqrt2 N\, \lambda^{1 / 2}(1, \tilde{m}_{K^{*}}^2, s_B)\, (C^{\nu_i\nu_j}_L+C^{\nu_i \nu_j}_R) \frac{V(s_B)}{1+\tilde{m}_{K^*}}, \nonumber \\[0.15cm]
  A^{ij}_{\|}(s_B) &= -2 \sqrt2 N\,(1+\tilde{m}_{K^*})\, (C^{\nu_i \nu_j}_L-C^{\nu_i \nu_j}_R)\, A_1(s_B), \nonumber \\[0.15cm]
  A^{ij}_{0}(s_B) &=-\frac{N(C^{\nu_i \nu_j}_L-C^{\nu_i \nu_j}_R)}{\widetilde{m}_{K^{*}} \sqrt{s_B}}\biggl[(1-\tilde{m}_{K^{*}}^2-s_B)(1+\tilde{m}_{K^{*}}) A_{1}(s_B) \nonumber \\
  &\hspace{6.5cm} -\lambda(1, \tilde{m}_{K^{*}}^2, s_B) \frac{A_2(s_B)}{1+\tilde{m}_{K^{*}}}\biggr],
\end{align}
with $\tilde{m}_{K^*}=m_{K^*} / m_B$ and $N= 1/2\,V_{tb}V_{ts}^*\left[G_F^2 \alpha_e^2 m_B^3s_B  \lambda^{1 / 2}(1, \tilde{m}_{K^{*}}^2, s_B)/(3\cdot 2^{10} \pi^5) \right]^{1/2}$. Similarly, the dineutrino invariant mass distribution of the $B \to K \nu \bar{\nu}$ decay can be written as~\cite{Colangelo:1996ay,Altmannshofer:2009ma,Buras:2014fpa}
\begin{align}
	\frac{d \Gamma(B \to K \nu \bar{\nu})}{d s_B} &= \frac{G_F^2 \alpha_e^2}{3\cdot 2^{10} \pi^5}|V_{tb}V_{ts}^*|^2 m_B^5 \lambda^{3/2}\left(1, \tilde{m}_{K}^2, s_B\right)\left[f_{+}^{B\to K}\left(s_B\right)\right]^2 \nonumber \\
	&\hspace{4.5cm} \times \sum_{i,j=e,\mu,\tau}\big\lvert C^{\nu_i\nu_j}_L+C^{\nu_i\nu_j}_R \big \rvert^2 \, .
\end{align}
For convenience, all the relevant form factors present in the above expressions are given in appendix~\ref{sec:form factor}. In addition, it is straightforward to obtain the distributions of other $b \to s (d) \nu \bar\nu$ decays, such as $B \to \pi \nu \bar\nu$ and $B_s \to \phi \nu \bar\nu$, from the above expressions with some replacements. Finally, one should notice that for charged $B$-meson decays, there is a long-distance contribution arising from the tree-level weak annihilation mediated by the on-shell $\tau$ lepton~\cite{Kamenik:2009kc}. For example, the $B^+ \to \tau^+(\to K^+ \bar{\nu}) \nu$ process provides an additional $\mO(10\%)$ contribution to the $B^+ \to K^+ \nu \bar\nu$ decay, and constitutes a background in the experimental extraction of the branching ratio~\cite{Belle-II:2023esi}. Here we do not include the tree-level contribution, which has already been subtracted away during the experimental data analysis~\cite{Belle-II:2023esi}.

\subsubsection{$b \to s \ell \ell$ decay}

The quark-level $b \to s \ell^+ \ell^-$ transitions, such as the rare $B_s \to \ell^+ \ell^-$, $B \to K^{(*)} \ell^+ \ell^-$ and $B_s \to \phi \ell^+ \ell^-$ decays, can provide promising probes of various NP effects (see refs.~\cite{London:2021lfn,Capdevila:2023yhq} for a recent review). For a general $b \to s \ell_i^- \ell_j^+$ transition, the effective weak Hamiltonian can be written as~\cite{Buchalla:1995vs}
\begin{align}\label{eq:Heff:bsmumu}
  \mathcal H_{\rm eff}^{b \to s \ell \ell}=-\frac{4G_F}{\sqrt{2}}  \frac{e^2}{16 \pi^2}  V_{tb}V_{ts}^*\sum_{k} C_k  \mO_k + \hc,
\end{align}
where the most relevant operators are $\mO_{9}^{(\prime)}$ and $\mO_{10}^{(\prime)}$. They are defined, respectively, by
\begin{align}\label{eq:operators:bsmumu}
  \mO_{9}^{ij} &= \big(\bar s \gamma_{\mu} P_L b \big) \big( \bar\ell_i \gamma^\mu \ell_j \big),
                &
  \mO_{10}^{ij} &= \big(\bar s  \gamma_{\mu} P_L b \big) \big( \bar\ell_i \gamma^\mu \gamma_5 \ell_j \big),
  \\[0.15cm]
  \mO_{9}^{\prime \, ij} &= \big(\bar s \gamma_{\mu} P_R b \big) \big( \bar\ell_i \gamma^\mu \ell_j \big),
                &
  \mO_{10}^{\prime \, ij} &= \big(\bar s  \gamma_{\mu} P_R b \big) \big( \bar\ell_i \gamma^\mu \gamma_5 \ell_j \big).
  \nonumber
\end{align}
In the SMEFT, the NP contributions to their Wilson coefficients are given by
\begin{align} 
  C_{9,\NP}^{ij} &= + c_\ell \big(\big{[}\mC_{lq}^{(1)}\big{]}_{ij23}+\big{[}\mC_{lq}^{(3)}\big{]}_{ij23}\big ),
  &
  C_{9,\NP}^{\prime \, i j}&= + c_\ell \big[\mC_{ld}\big]_{ij 23},
  \\ 
  C_{10,\NP}^{ij} &= - c_\ell \big( \big{[}\mC_{lq}^{(1)}\big{]}_{ij23} + \big{[}\mC_{lq}^{(3)}\big{]}_{ij23}\big ),
  &
  C_{10,\NP}^{\prime \, i j}  & = - c_\ell \big[\mC_{ld}\big]_{ij 23},
                                 \nonumber
\end{align}
with the normalization constant $c_\ell \equiv (\pi /(\alpha_e V_{tb}V_{ts}^*)) \cdot (v^2/\Lambda^2)$. In the MFV framework, from eq.~(\ref{eq:MFV:SMEFT}), we can further obtain
\begin{align}
  C_{9,\MFV}^{ij}&= - C_{10,\MFV}^{ij}=  c_\ell \Delta_q^{23}  \big[ \epsilon_1^{(1)} \big( \kappa_0^{(1)} \delta^{ij} + \kappa_1^{(1)} \Delta_\ell^{ij} \big) + \big( (1)\to (3) \big) \big], \nonumber \\
  C_{9,\MFV}^{\prime \, ij}&= C_{10,\MFV}^{\prime \, i j}   = 0,
\end{align}
where $\epsilon_{1}^{(n)}$ and $\kappa_{0,1}^{(n)}$ denote the MFV parameters for the operator $\mQ_{lq}^{(n)}$ ($n=1,3$). For the $B\to \pi \ell_i \ell_j$, $B \to K^{(*)} \ell_i \ell_j$ and $B_s \to \ell_i \ell_j$ decays, which are all induced by the quark-level $b \to s(d) \ell_i \ell_j$ transitions and will be included in our numerical analysis. These decays have been studied in refs.~\cite{Gershtein:1976mv,Khlopov:1978id,Crivellin:2015era,Lee:2015qra,Zhang:2018nmy} and the explicit expressions of their branching ratios can be found in, e.g., refs.~\cite{Crivellin:2015era,Lee:2015qra,Zhang:2018nmy}.

\subsubsection{$s \to d \nu \bar{\nu}$ decay}

Both the $K^+ \to \pi^+ \nu \bar{\nu}$ and $K_L \to \pi^0 \nu \bar{\nu}$ decays are induced by the $s \to d \nu_i \bar{\nu_j}$ transition. The corresponding effective weak Hamiltonian, including the effective operators and their Wilson coefficients, can be obtained from the ones of the $b \to s \nu \bar\nu$ transition in section~\ref{sec:b2snunu} by replacing the related flavour indices. In addition, for the $K^+ \to \pi^+ \nu \bar{\nu}$ decay, a non-negligible charm contribution must be taken into account~\cite{Buchalla:1998ba,Misiak:1999yg}.

In absence of the operator $O_R^{\nu_i \nu_j}$, the branching ratio of the $K^+ \to \pi^+ \nu \bar\nu$ decay is given by~\cite{Buras:2004uu,Mescia:2007kn,Cirigliano:2011ny} 
\begin{align}\label{eq:br_K2pinunu}
  \mB (K^+ \to \pi^+ \nu \bar\nu ) &= \kappa_+(1+\Delta_{\rm EM}) \sum_{i,j=e,\mu,\tau} \frac{1}{3}\bigg[\bigg(\sin^2\theta_W\frac{{\rm Im} [\lambda_t C_L^{\nu_i \nu_j}]}{2\lambda^5} \bigg)^2 \nonumber \\
  & \hspace{2.0cm} + \bigg(\frac{{\rm Re} \lambda_c}{\lambda} P_c(X)\delta^{ij} - \sin^2\theta_W\frac{{\rm Re} [\lambda_t C_L^{\nu_i \nu_j}]}{2\lambda^5}\bigg)^2\,\bigg]\,,
\end{align}
where $\lambda_i = V_{is}^*V_{id}$, $\lambda = |V_{us}|$, and $\sin^2\theta_W=0.23116$ is the weak mixing angle. The parameter $\kappa_+ = (5.173 \pm 0.025)\times 10^{-11} (\lambda/0.225)^8$ contains, besides some other factors, the hadronic matrix element of the weak current $\bar{s}\gamma_\mu P_L d$ that can be extracted from the $K_{\ell 3}$ data with the help of isospin symmetry~\cite{Buras:2015qea}. $\Delta_{\rm EM} = -0.003$ accounts for the isospin-breaking electromagnetic  corrections~\cite{Mescia:2007kn,Cirigliano:2011ny}. $P_c(X) = 0.404\pm 0.024$ comprises the short- and long-distance charm contributions~\cite{Buras:2015qea,Buras:2006gb,Brod:2008ss,Isidori:2005xm}.

The rare decay $K_L \to \pi^0 \nu \bar\nu $ proceeds almost entirely through direct CP violation in the SM~\cite{Littenberg:1989ix,Buchalla:1998ux}, and the charm contribution can be fully neglected~\cite{Buras:2004uu}. The branching ratio of this decay can be written as~\cite{Buras:2004uu,Buchalla:1996fp}
\begin{align}\label{eq:br_KL2pinunu}
	\mB (K_L \to \pi^0 \nu \bar\nu )= \kappa_L (1-\delta_\epsilon) \sum_{i,j=e,\mu,\tau} \frac{1}{3}\left(\frac{\sin^2\theta_W}{2\lambda^5} \text{Im} \big[\lambda_t C_L^{\nu_i \nu_j}\big]\right)^2,
\end{align} 
where $\kappa_L = (2.231\pm 0.013)\times 10^{-10}(\lambda/0.225)^8$ encodes, besides some other factors, the hadronic matrix element that can be again extracted from the $K_{\ell 3}$ decay with the help of isospin symmetry~\cite{Marciano:1996wy,Mescia:2007kn}. The parameter $\delta_\epsilon$ encodes, on the other hand, the highly suppressed indirect CP-violating contribution~\cite{Buchalla:1998ux}.

\section{$\boldsymbol{M_1 \to M_2 + \text{DM}}$}
\label{sec:theory:DM}

In this section, we first recapitulate the theoretical framework of the DSMEFT and DLEFT, and then detail the calculation of $M_1 \to M_2 + \text{DM}$ decays.

\subsection{DSMEFT}
\label{sec:DSMEFT}

It is known that EFT can provide a systematic framework for parameterizing various effects induced by the DM particles. The DSMEFT is a generalization of the SMEFT~\cite{Buchmuller:1985jz,Grzadkowski:2010es} to include the DM particles, which can be either lighter or heavier than the electroweak scale and are gauge singlets under the SM gauge group~\cite{Aebischer:2022wnl,Liang:2023yta,Song:2023jqm}. It is also assumed that electroweak symmetry breaking must be implemented via the usual Higgs mechanism. Following the convention of ref.~\cite{Aebischer:2022wnl}, the DSMEFT Lagrangian takes the form
\begin{align}\label{eq:DSMEFT:Lagrangian}
  \mL_{\mathrm{DSMEFT}} &\supset \frac{1}{\Lambda} \sum_{i} \mC_i \mQ_i^{(5)} +  \frac{1}{\Lambda^2} \sum_j \mC_j \mQ_j^{(6)},
\end{align}
where $\Lambda$ is the NP scale above which new degrees of freedom appear and interact with the SM and DM particles. The dim-$d$ effective operators $\mQ_i^{(d)}$ involve at least one DM field and respect the SM $SU(3)_C \otimes SU(2)_L \otimes U(1)_Y$ gauge symmetry. $\mC_i$ denote the corresponding dimensionless Wilson coefficients. We will define the DSMEFT at the electroweak scale $\mu_{\rm EW}$.

In the DSMEFT framework, three types of DM particles can be added: spin-0 scalar $\phi$, spin-1/2 fermion $\chi$ and spin-1 vector $X$. Generally, each type of the DM particles can have multiple generations. For simplicity, the case of one generation DM will be assumed in this paper. For the operators vanishing in the case of one generation, however, two generations of DM particles will be considered. Furthermore, in order to affect the rare $B$-meson and kaon FCNC decays, only the operators that can induce tree-level FCNC interactions among the down-type quarks are included in our analysis. Finally, motivated by the large excess of $\mB(B^+ \to K^+ + \nu \bar\nu)$ reported by Belle II~\cite{Belle-II:2023esi}, we will focus on the dim-5 and dim-6 operators, which are less suppressed than the ones with dim $\geq 7$. Consequently, the DSMEFT operators we are considering must involve two down-type quarks and at least one DM field. Furthermore, in order to mimic the rare $B$-meson and kaon FCNC decays involving a dineutrino final state, the DM particles must be light enough to be produced in these decays. For each type of the DM particles, the relevant effective operators are given in the following.

For the scalar DM particles, there are four operators,
\begin{align}\label{eq:DSMEFT:S}
  \mQ_{d\phi} &= \big( \bar{q}_p d_r H \big) \phi + \hc ,
  &
  \mQ_{d\phi^2} &= \big( \bar{q}_p d_r H \big) \phi^2 + \hc , \nonumber \\
  \mQ_{\phi q}  &= \big( \bar{q}_p \gamma_\mu q_r \big)\big (i\phi_1 \overleftrightarrow{\partial^\mu}\phi_2\big) ,
  &
  \mQ_{\phi d} &= \big( \bar{d}_p \gamma_\mu d_r \big)\big(i\phi_1 \overleftrightarrow{\partial^\mu}\phi_2\big),
\end{align}
where $\phi_1 \overleftrightarrow{\partial^\mu}\phi_2 = \phi_1 ({\partial^\mu}\phi_2) - ({\partial^\mu}\phi_1) \phi_2$, and $\phi_{(i)}$ is the real scalar DM field with the generation index characterized by $i$. For the operators which are not hermitian, ``$+\hc$'' should be added to them to indicate that the hermitian conjugated operators should be included in the DSMEFT Lagrangian eq.~(\ref{eq:DSMEFT:Lagrangian}). In addition, constructions of the operators $\mQ_{\phi q}$ and $\mQ_{\phi d}$ requires at least two generations of scalar DM fields.

For the fermionic DM particles, there are only two operators, 
\begin{align}\label{eq:DSMEFT:F}
  \mQ_{q\chi} = \big(\bar{q}_{p} \gamma_\mu q_{r} \big) (\bar{\chi} \gamma^\mu \chi \big),
  \qquad \qquad 
  \mQ_{d\chi} = \big(\bar{d}_{p} \gamma_\mu d_{r} \big) (\bar{\chi} \gamma^\mu \chi \big),
\end{align}
where $\chi$ denotes the fermionic DM field. In this paper, all the observables considered are insensitive to the chirality of the fermionic DM. Without loss of generality, $\chi$ will be therefore chosen as the right-handed Dirac field.

For the vector DM particles, by using the field strength tensor as the building block, there is only one operator,
\begin{align}\label{eq:DSMEFT:V:1}
  \mQ_{dHX} = \big( \bar{q}_{p}\sigma_{\mu\nu} d_{r} \big)H X^{\mu\nu} + \hc,
\end{align}
where $X_{\mu\nu}= \partial_\mu X_\nu - \partial_\nu X_\mu$, and $X_\mu$ denotes the real vector DM field. Taking $X_\mu$ an additional building block, the following operators also appear~\cite{Song:2023jqm}:
\begin{align}\label{eq:DSMEFT:V:2}
  \mQ_{dX} &= \big( \bar d_p \gamma_\mu d_r \big) X^\mu,
  &
    \mQ_{H d X} &= \big( H^\dagger H \big) \big( \bar d_p \gamma^\mu d_r \big)X_\mu,
  \nonumber\\
  \mQ_{qX} &= \big( \bar q_p \gamma_\mu q_r \big) X^\mu,
  &
    \mQ_{HqX}^{(1)} &= \big( H^\dagger H \big) \big( \bar q_p \gamma^\mu q_r \big)X_\mu,
  \nonumber\\
  \mQ_{dX^2} & =  \big( \bar q_p d_r H \big) X_\mu X^\mu + \hc ,
  &
  \mQ_{HqX}^{(3)} &= \big( H^\dagger \tau^I H \big) \big( \bar{q}_p \tau^I  \gamma^\mu  q_r \big)  X_\mu,
  \nonumber\\
  \mQ_{qXX} &=   \big( \bar q_p \gamma_\mu q_r \big) X^{\mu \nu}X_\nu\,,
  &
    \mQ_{dXX} &= \big( \bar d_p \gamma_\mu d_r \big) X^{\mu \nu}X_\nu, 
  \nonumber\\
  \mQ_{q\tilde X X} & = \big( \bar q_p \gamma_\mu q_r \big) \tilde X^{\mu\nu}X_\nu \,,
  &
    \mQ_{d\tilde X X} & = \big( \bar d_p \gamma_\mu d_r \big) \tilde X^{\mu\nu}X_\nu \,,
  \nonumber\\
  \mQ_{DqX^2} &=  i\big( \bar q_p \gamma^\mu D^\nu q_r \big) X_\mu X_\nu + \hc,
  &
  \mQ_{DdX^2} & =  i\big( \bar d_p  \gamma^\mu D^\nu d_r \big)X_\mu X_\nu + \hc,
  \nonumber \\
  \mQ_{dHX^2} & = (\bar q_p \sigma_{\mu\nu} d_r H) X_1^\mu X_2^\nu  + \hc,
\end{align}
where $\tilde X^{\mu\nu}= \epsilon^{\mu\nu\rho\sigma}X_{\rho\sigma}/2$. It is noted that the rare $B$-meson and kaon decay rates induced by the operators involving the vector DM field $X_\mu$ are divergent in the $m_X \to 0$ limit. Such a divergence is caused by the longitudinal polarization of the vector field and related to the phenomenon that it is impossible to consistently define a massless limit for a vector field without an active gauge symmetry~\cite{Kamenik:2011vy}. As in refs.~\cite{Kamenik:2011vy,Williams:2011qb,He:2022ljo}, this singularity is treated by assuming some kind of Higgs mechanism in the dark sector. As a result, the Wilson coefficients of the operators in eq.~(\ref{eq:DSMEFT:V:2}) should be proportional to some powers of $(m_X /\Lambda)$ and can be redefined as
\begin{align}
  \mC_i = \tilde \mC_i \cdot
  \begin{cases}
    (m_X /\Lambda )^2 & \text{ for } \mQ_i =  \mQ_{dX^2}, \mQ_{DdX^2}, \mQ_{DqX^2}, \mQ_{dHX^2},
    \\
    (m_X /\Lambda ) & \text{ for } \mQ_i = \text{others}.
  \end{cases}
\end{align}
These redefined Wilson coefficients $\tilde\mC_i$ will be used in the following analysis.

We can also include the axion or axion-like particles (ALPs)~\cite{Peccei:1977hh,Peccei:1977ur,Weinberg:1977ma,Wilczek:1977pj} in the EFT approach~\cite{Georgi:1986df,Brivio:2017ije,Chala:2020wvs,Bauer:2020jbp,Galda:2021hbr,Song:2023lxf}. They are pseudo Nambu-Goldstone bosons with non-trivial properties, and can emerge from spontaneous breaking of some global symmetries. Up to dim-5, there are only two operators,
\begin{align}
  \mQ_{qa} = \big( \bar q_p \gamma_\mu q_r \big) \partial^\mu a,
  \qquad \qquad
  \mQ_{da} = \big( \bar d_p \gamma_\mu d_r \big) \partial^\mu a,
\end{align}
where $a$ denotes the ALP field.

In the MFV hypothesis discussed in section~\ref{sec:MFV}, the Wilson coefficients of the DSMEFT operators should take special flavour structures. Explicitly, by using the quark currents in eqs.~\eqref{eq:MFV:vector current:dL}-\eqref{eq:MFV:vector current:dR}, we obtain 
\begin{align}\label{eq:WC:MFV}
  \mC^\MFV_i =
  \begin{cases}
    \epsilon_0^i \hat{\lambda}_d + \epsilon_1^i \Delta_q \hat{\lambda}_d
    & \text{for } \mQ_i = \mQ_{d\phi}, \mQ_{d\phi^2}, \mQ_{dHX}, \mQ_{dHX^2}, \mQ_{dX^2},
    \\
    \epsilon_0^i \mathds{1} + \epsilon_1^i \Delta_q
    & \text{for } \mQ_i = \mQ_{\phi q}, \mQ_{q\chi}, \mQ_{qXX},  \mQ_{q \tilde X X},  \mQ_{DqX^2} , \mQ_{qX}, \mQ_{HqX}^{(1,3)}, \mQ_{qa},
    \\
    \epsilon_0^i \mathds{1}
    & \text{for } \mQ_i= \mQ_{\phi d}, \mQ_{d\chi}, \mQ_{dXX},  \mQ_{d \tilde X X}, \mQ_{DdX^2} , \mQ_{dX}, \mQ_{HdX}, \mQ_{da},
  \end{cases}
\end{align}
for the DSMEFT operators discussed above. Here the Wilson coefficient $\mC^\MFV_i$ are given for the down-type quark currents expressed in the mass eigenbasis, and they are $3 \times 3$ matrices in flavour space. Generally, the real parameters $\epsilon_{0,1}^i$ are different from each other for different operators $\mQ_{i}$. We can see that the operators listed in the last line of the above equation do not generate tree-level FCNC interactions among the down-type quarks.

\subsection{DLEFT}
\label{eq:DLEFT}

Below the electroweak scale, the NP effects induced by the DM particles can be described by the DLEFT~\cite{Aebischer:2022wnl}. This EFT is obtained by integrating out all the heavy particles with masses at or above the electroweak scale and can be considered as an extension of the LEFT~\cite{Buchalla:1995vs,Jenkins:2017jig} by including besides the light SM fields also the light DM particles. The DLEFT respects the SM $SU(3)_C\otimes U(1)_{\rm em}$ gauge symmetry, and does not contain the Higgs boson. Thus, it can be considered without reference to the DSMEFT for the light DM and light SM particles interacting at energies below the electroweak scale. 
Following the same convention as in ref.~\cite{Aebischer:2022wnl}, we can write the DLEFT Lagrangian as
\begin{align}
  \mL_{\mathrm{DLEFT}} \supset \frac{1}{\Lambda} \sum_{i} L_i \mO_i^{(5)} +  \frac{1}{\Lambda^2} \sum_j L_j \mO_j^{(6)},
\end{align}
up to dim-6, where $\mO_i^{(d)}$ denote the dim-$d$ effective operators and $L_i$ the corresponding Wilson coefficients.

Analogous to the DSMEFT case, we can write down all the relevant DLEFT operators involving the scalar, fermionic and vector DM fields. For the scalar DM, we have
\begin{align}\label{eq:DLEFT:S}
  \mO_{d\phi} &= (\bar{d}_{Lp} d_{Rr})\phi + \hc,
  &
  \mO_{\phi d}^L &= (\bar{d}_{Lp} \gamma_\mu d_{Lr}) \big(i\phi_1 \overleftrightarrow{\partial^\mu}\phi_2\big),     
  \nonumber\\
  \mO_{d\phi^2} &= (\bar{d}_{Lp} d_{Rr})\phi^2 + \hc,
  &
  \mO_{\phi d}^R &= (\bar{d}_{Rp} \gamma_\mu d_{Rr}) \big(i\phi_1 \overleftrightarrow{\partial^\mu}\phi_2\big),
\end{align}
where $d_L$ and $d_R$ denote the left-handed and right-handed down-type quarks, respectively. For the fermionic DM, the relevant operators are given by
\begin{align}\label{eq:DLEFT:F}
  \mO_{d\chi}^{V,\, LR} &= (\bar{d}_{Lp} \gamma_\mu d_{Lr}) (\bar{\chi}_a \gamma^\mu \chi_b),
  &
  \mO_{d\chi}^{V,RR} &= (\bar{d}_{Rp} \gamma_\mu d_{Rr}) (\bar{\chi}_a \gamma^\mu \chi_b),
  \nonumber \\
  \mO_{d\chi}^{S,LR} &= (\bar{d}_{Rp} d_{Lr}) (\chi_a^T C \chi_b) + \hc,
  &
  \mO_{d\chi}^{S,RR} &=  (\bar{d}_{Lp} d_{Rr}) (\chi_a^T C \chi_b) + \hc,
  \nonumber \\
  \mO_{d\chi}^{T,RR} &= (\bar{d}_{Lp} \sigma^{\mu\nu} d_{Rr}) (\chi_a^T C \sigma_{\mu\nu} \chi_b) + \hc,
\end{align}
with the charge conjugate operator defined by $C=i \gamma^2 \gamma^0$. For the vector DM, there is only one operator,
\begin{align}\label{eq:DLEFT:V:1}
  \mO_{dX}^T &= (\bar{d}_{Lp}\sigma_{\mu\nu} d_{Rr}) X_a^{\mu\nu} + \hc,
\end{align}
by using the field strength $X_{\mu\nu}$ as the building block. However, including $X_\mu$ as additional building block, the following operators are obtained:
\begin{align}\label{eq:DLEFT:V:2}
  \mO_{dX}^L &= \big( \bar d_{Lp} \gamma_\mu d_{Lr} \big) X^\mu ,
  &
  \mO_{dX}^R &= \big( \bar d_{Rp} \gamma_\mu d_{Rr} \big) X^\mu,
  \nonumber\\
  \mO_{dXX}^L &=   \big( \bar d_{Lp} \gamma_\mu d_{Lr} \big) X^{\mu \nu}X_\nu,
  &
  \mO_{dXX}^R &=   \big( \bar d_{Rp} \gamma_\mu d_{Rr} \big) X^{\mu \nu}X_\nu,
  \nonumber\\
  \mO_{d\tilde X X}^L & = \big( \bar d_{Lp} \gamma_\mu d_{Lr} \big) \tilde X^{\mu\nu}X_\nu,
  &
  \mO_{d\tilde X X}^R & = \big( \bar d_{Rp} \gamma_\mu d_{Rr} \big) \tilde X^{\mu\nu}X_\nu ,
  \nonumber\\
  \mO_{DdX^2}^L &=  i\big(\bar d_{Lp} \gamma^\nu  D^\mu d_{Lr} \big) X_\mu X_\nu + \hc ,
  &
  \mO_{DdX^2}^R &=  i\big( \bar d_{Rp} \gamma^\nu  D^\mu d_{Rr} \big) X_\mu X_\nu + \hc ,
  \nonumber\\
  \mO_{dX^2} & = \big( \bar d_{Lp} d_{Rr} \big) X_\mu X^\mu +\hc,
  &
  \mO_{dX^2}^T & = \big( \bar d_{Lp} \sigma_{\mu\nu} d_{Rr} \big) X_1^\mu X_2^\nu +\hc,
\end{align}
where $D_\mu = \partial_\mu + i g_s T^a G_\mu^a + ie Q A_\mu$ denotes the covariant derivative of the $SU(3)_C \otimes U(1)_{\rm em}$ gauge symmetry. Here we have chosen the basis where the combinations of $X_{\mu\nu}$ and $X_\mu$ are similar as in the DSMEFT operators of eq.~(\ref{eq:DSMEFT:V:2}). One can prove that this basis is equivalent to the one adopted in ref.~\cite{He:2022ljo}. In addition, there are two operators involving the ALP field, which take the form
\begin{align}
  \mO_{da}^L = \big( \bar d_{Lp} \gamma_\mu d_{Lr} \big) \partial^\mu a,
  \qquad\qquad
  \mO_{da}^R = \big( \bar d_{Rp} \gamma_\mu d_{Rr} \big) \partial^\mu a.
\end{align}

Although the DLEFT can be considered without reference to the DSMEFT, the former should arise as the low-energy limit of the latter, if we start from the DSMEFT. At the electroweak scale $\mu_{\rm EW}$, the tree-level matching conditions between these two EFTs are summarized below:
\begin{itemize}
  \item Scalar DM:
\begin{align}
  L_{\phi d}^L &= \mC_{\phi q},
  &
  L_{\phi d}^R &= \mC_{\phi d},
  &
  L_{d\phi^2} &= \frac{v}{\sqrt{2}\Lambda} \mC_{d\phi^2},
  &
  L_{d\phi} &= \frac{v}{\sqrt{2}\Lambda} \mC_{d\phi}.
\end{align}
\item Fermionic DM:
\begin{align}
  L_{d \chi}^{V,\, LR} &= \mC_{q\chi},
  &
    L_{d \chi}^{V,\, RR} &= \mC_{d\chi},
  &
    L_{d \chi}^{S,\, LR} &= 0,
  &
    L_{d \chi}^{S,\, RR} &= 0,
  &
    L_{d \chi}^{T,\, RR} &= 0.
\end{align}
\item Vector DM:
\begin{align}
  L_{dX}^T = \frac{v}{\sqrt{2}\Lambda} \mC_{dHX},
  L_{dX}^L = \mC_{qX} + \frac{v^2}{2 \Lambda^2} \big(\mC_{Hq X}^{(1)} + \mC_{HqX}^{(3)}\big) ,
  L_{dX}^R = \mC_{dX} + \frac{v^2}{2 \Lambda^2} \mC_{HdX},
  \nonumber
\end{align}
\begin{align}
  L_{dXX}^L &= \mC_{qXX},
  &
    L_{d\tilde X X}^L & = \mC_{q\tilde X X} ,
  &
    L_{DdX^2}^L &= \mC_{DqX^2},
  &
    L_{dX^2} & =  \frac{v}{\sqrt{2}\Lambda} \mC_{dX^2},
      \nonumber\\
  L_{dXX}^R &= \mC_{dXX},
  &
    L_{d\tilde X X}^R & = \mC_{d\tilde X X},
  &
    L_{DdX^2}^R &= \mC_{DdX^2},
  &
    L_{dX^2}^T & =  \frac{v}{\sqrt{2}\Lambda} \mC_{dHX^2}.
\end{align}
\item ALP:
\begin{align}
  L_{da}^L = \mC_{qa},
  \qquad \qquad
  L_{da}^R = \mC_{da}.
\end{align}
\end{itemize}
Here all the Wilson coefficients are given in the mass eigenbasis. In the MFV hypothesis, the Wilson coefficients in DSMEFT take the form of eq.~(\ref{eq:WC:MFV}).

Since the above matching conditions are given at the electroweak scale, i.e., $L_i \equiv L_i(\mu_{\rm EW})$ and $\mC_i \equiv \mC_i(\mu_{\rm EW})$, the renormalization group (RG) evolution should be performed in the DLEFT to obtain the corresponding Wilson coefficients at the typical scales of $B$-meson and kaon decays. In the DLEFT, as the DM particles are color singlet, we need only consider the QCD RG evolution of the quark current in each DLEFT operator. Furthermore, the different DLEFT operators do not mix under the QCD RG running. Accordingly, the RG equations take the form
\begin{align}\label{eq:RGE}
  \frac{d L_i (\mu)}{d \ln\mu} = \frac{\alpha_s}{4\pi}\gamma_i \, L_i(\mu),
\end{align}
where $L_i(\mu)$ denote the Wilson coefficients of the operators $\mO_i$ at the scale $\mu$. $\gamma_i$ denote the anomalous dimension matrices (ADM) for the operators $\mO_i$. They can be obtained from the ADM of the semi-leptonic operators in the LEFT~\cite{Buchalla:1995vs,Jenkins:2017dyc}. For the DLEFT operators with a non-zero matching condition, the values of these ADM are given by
\begin{align}
  \gamma_i =
  \begin{cases}
    -6 C_F & \text{for } \mO_i= \mO_{d\phi},\mO_{d\phi^2}, \mO_{dX^2},
    \\
    +2 C_F& \text{for } \mO_i = \mO_{dX}^T, \mO_{dX^2}^T,
    \\
    0 & \text{for } \mO_i = \mO_{\phi d}^{L,R}, \mO_{d\chi}^{V,LR}, \mO_{d\chi}^{V,RR}, \mO_{dX}^{L,R}, \mO_{dXX}^{L,R}, \, \mO_{d \tilde X X}^{L,R}, \mO_{DdX^2}^{L,R},
  \end{cases}
\end{align}
with $C_F = (N_c^2-1)/(2N_c)$ and $N_c=3$. Numerically, we find
\begin{align}
  L_i(4.8\GeV) &= L_i(m_Z) \cdot
  \begin{cases}
    1.37  & \text{for } \mO_i= \mO_{d\phi},\mO_{d\phi^2}, \mO_{dX^2},
    \\
    0.90  & \text{for } \mO_i = \mO_{dX}^T, \mO_{dX^2}^T,
  \end{cases}
    \\  
  L_i(2.0\GeV) &=  L_i(m_Z) \cdot
  \begin{cases}
    1.59 & \text{for } \mO_i= \mO_{d\phi},\mO_{d\phi^2}, \mO_{dX^2},
    \\
    0.86 & \text{for } \mO_i = \mO_{dX}^T, \mO_{dX^2}^T,
  \end{cases}
\end{align}
which can be used in the $B$-meson and kaon decays, respectively.

\subsection{Observables}
\label{sec:obs:DM}

Within the DLEFT, by using the similar treatment as for the $d_i \to d_j \nu \bar\nu$ transitions discussed in section~\ref{sec:obs:nu}, one can calculate the various $d_i \to d_j + \text{DM}$ processes. Generally, the amplitudes of these decays can be factorized into the product of the Wilson coefficient and the hadronic matrix element of the corresponding effective operator that can be parameterized form factors or decay constants. In appendix~\ref{eq:hadronic matrix element}, we summarize the relevant hadronic matrix elements used in this paper. For the explicit expressions of the decay rates, we refer the readers to refs.~\cite{Bird:2004ts,Li:2019cbk,Kamenik:2011vy,He:2020jly,Geng:2020seh,Li:2021sqe,Kling:2022uzy} for the $d_i \to d_j  +\phi\, (+\phi)$ decays, refs.~\cite{Su:2019tjn,Kamenik:2011vy,Li:2020dpc,Felkl:2021uxi} for the $d_i \to d_j + \chi + \chi$ decay, refs.~\cite{Kamenik:2011vy,Li:2021sqe,He:2022ljo} for the  $d_i \to d_j + X \,(+X)$ decays, and refs.~\cite{MartinCamalich:2020dfe,Bauer:2021mvw,Guerrera:2022ykl} for the $d_i \to d_i + a$ decays. To the best of our knowledge, a complete computation of the longitudinal polarization fraction $F_L$ of the vector meson $V$ in $P \to V + \DM$ decays within the DLEFT is still missing, although the cases with some effective operators have been studied previously, e.g., in refs.~\cite{Altmannshofer:2009ma,Kamenik:2011vy}. In this work, we systematically calculate the polarization observables in the DLEFT. All the relevant analytical expressions have been implemented in our recently developed package \texttt{HadronToNP}~\cite{HadronToNP:2024}, which will be used in the following numerical analysis.

\section{Numerical analysis}
\label{sec:numerical analysis}

In this section, we proceed to present our numerical results and discussions. In table~\ref{tab:input}, we list the main input parameters used in our numerical analysis. Table~\ref{table:observables} summaries our SM predictions and the up-to-date experimental measurements of the relevant observables. We take $\Lambda_\NP = 1\TeV$ in the SMEFT and DSMEFT. In the following, we investigate two explanations to the recent Belle II measurement of $\mB(B^+ \to K^+ \nu \bar\nu)$~\cite{Belle-II:2023esi}. In section~\ref{sec:num:bsnunu}, we consider the $b \to s + \nu \bar\nu$ transition in the SMEFT. In section~\ref{sec:num:bsDM}, the missing energy signals in the Belle II measurement are interpreted in the DSMEFT as contributions from DM particles.

\begin{table}[t]
  \renewcommand*{\arraystretch}{1.2}
  \centering
  \begin{tabular}{c | c | c | c}
    \toprule
    Input & Value & Unit &  Reference
    \\\midrule
    $m_t^{\text{pole}}$ & $172.69 \pm 0.30$&  GeV&  \cite{Workman:2022ynf}
    \\
    $|V_{cb}|$(semi-leptonic) & $41.15 \pm 0.34 \pm 0.45 $ & $10^{-3}$ &  \cite{CKMfitter}
    \\
    $|V_{ub}|$(semi-leptonic) & $3.88 \pm 0.08 \pm 0.21  $ & $10^{-3}$ &  \cite{CKMfitter}
    \\
    $|V_{us}|f^{K \to \pi}_+(0)$ & $0.2165 \pm 0.0004 $ &&  \cite{CKMfitter}
    \\
    $\gamma$&  $ 72.1^{+5.4}_{-5.7}  $ & $^\circ$ &  \cite{CKMfitter}
    \\
    $f^{K \to \pi}_+(0)$  & $0.9675 \pm 0.0009 \pm 0.0023$ &&   \cite{CKMfitter}
    \\\midrule
    $ \Delta m_{21}^2 $ & $7.41^{+0.21}_{-0.20} $ ($7.41^{+0.21}_{-0.20} $ ) & $10^{-5}\cdot \text{eV}^2$ &  \cite{Workman:2022ynf}
    \\
    $ \Delta m_{32}^2 $ & $2.437^{+0.028}_{-0.027}$ ($-2.498^{+0.032}_{-0.025}$) & $10^{-3} \cdot \text{eV}^2$ &  \cite{Workman:2022ynf}
    \\
    $\theta_{12}$ &  $33.41^{+0.75}_{-0.72} $ ($33.41^{+0.75}_{-0.72} $) & $^\circ$ & \cite{Workman:2022ynf}
    \\
    $\theta_{23}$ &  $49.1^{+1.0}_{-1.3} $ ($49.5^{+0.9}_{-1.2} $) & $^\circ$ &  \cite{Workman:2022ynf} \\
    $\theta_{13}$ &  $8.54^{+0.11}_{-0.12} $ ($8.57^{+0.12}_{-0.11} $) & $^\circ$ &  \cite{Workman:2022ynf}
    \\
    $\delta_{\rm CP}$ &  $197^{+42}_{-25} $ ($286^{+27}_{-32} $) & $^\circ$ & \cite{Workman:2022ynf}
    \\\bottomrule
  \end{tabular}
  \caption{Main input parameters used in our numerical analysis. The values (in brackets) of the parameters in the lepton sector correspond to the NO (IO) of neutrino masses.}
  \label{tab:input}
\end{table}

\begin{table}[t]
  \renewcommand*{\arraystretch}{1.2}
  \centering
  \begin{tabular}{c| c |c|c|c}
    \toprule
    Observable    &  SM   & Exp & Unit & Sensitivity  \\
    \midrule
    $\mB(B^+ \to K^+ \nu \bar{\nu}) $     & $4.16 \pm 0.57$ & $23 \pm 5 ^{+5}_{-4}$~\cite{belleIItalk2023}       &  $10^{-6} $  & 19\%~\cite{Belle-II:2022cgf} \\
    $\mB(B^0 \to K^0 \nu \bar{\nu})$      & $3.85 \pm 0.52$ & $<26 $~\cite{Belle:2017oht}                       &  $ 10^{-6}$  & 87\%~\cite{Belle-II:2022cgf}\\
    $\mB(B^+ \to K^{*+} \nu \bar{\nu})$   & $9.70 \pm 0.94$ & $<61 $~\cite{Belle:2017oht}                       &  $ 10^{-6} $  & 75\%~\cite{Belle-II:2022cgf} \\
    $\mB(B^0 \to K^{*0} \nu \bar{\nu}) $  & $9.00 \pm 0.87$ & $<18 $~\cite{Belle:2017oht}                       &  $ 10^{-6}$   & 40\%~\cite{Belle-II:2022cgf}\\
    $\mB(B_s \to \phi \nu \bar\nu)$      & $9.93 \pm 0.72$  & $<5400$~\cite{DELPHI:1996ohp}    &  $ 10^{-6}$   & \;\;2\%~\cite{Li:2022tov}\\
    \midrule
    $\mB(B^+ \to \pi^+ \nu \bar{\nu}) $   & $1.40 \pm 0.18$ & $<140$~\cite{Belle:2017oht}                       & $ 10^{-7} $   & \\
    $\mB(B^0 \to \pi^0 \nu \bar{\nu}) $   & $6.52 \pm 0.85$ & $<900$~\cite{Belle:2017oht}                       & $ 10^{-8} $   & \\
    $\mB(B^+ \to \rho^+ \nu \bar{\nu}) $  & $4.06 \pm 0.79$ & $<300$~\cite{Belle:2017oht}                       & $ 10^{-7} $   & \\
    $\mB(B^0 \to \rho^0 \nu \bar{\nu}) $  & $1.89 \pm 0.36$ & $<400$~\cite{Belle:2017oht}                       & $ 10^{-7} $   &  \\
    \midrule
    $\mB(K^+ \to  \pi^+ \nu \bar{\nu}) $  & $8.42 \pm 0.61$ & $10.6 ^{+4.0}_{-3.4} \pm0.9 $~\cite{NA62:2021zjw}   & $ 10^{-11} $  &   \\
    $\mB(K_L \to  \pi^0 \nu \bar{\nu}) $  & $3.41 \pm 0.45$ & $<300 $~\cite{KOTO:2018dsc}                       & $ 10^{-11} $   &  \\
    \midrule
    $\mB(B_s \to \nu \bar\nu)$            & $\approx 0$     & $<5.9$~\cite{Alonso-Alvarez:2023mgc}               & $10^{-4}$    & $<1.1 \times 10^{-5}$~\cite{Belle-II:2018jsg} \\
    $\mB(B^0 \to \nu \bar\nu)$            & $\approx 0$     & $<1.4$~\cite{Alonso-Alvarez:2023mgc}               & $10^{-4}$    & $<5.0 \times 10^{-6}$~\cite{Belle-II:2018jsg} \\
    \bottomrule
  \end{tabular}
  \caption{Our SM predictions and the experimental measurements of the relevant observables. Upper limits are all given at $90\%$ confidence level (CL). The last column shows the expected sensitivities at the Belle II ($5\ab^{-1}$) or at the CEPC (Tera-$Z$ phase). The sensitivities at the FCC-ee are very similar to that at the CEPC, and hence are not given here.}
  \label{table:observables}
\end{table}

\subsection[$M_1 \to M_2 + \nu \bar\nu$]{$\boldsymbol{M_1 \to M_2 + \nu \bar\nu}$}
\label{sec:num:bsnunu}

The recent Belle II measurement of the $B^+ \to K^+ + \inv$ decay could be explained by NP contributions to the $b \to s \nu \bar\nu$ transition. In this case, the theoretical prediction of the related $M_1 \to M_2 + \inv$ processes takes the form
\begin{align}
  \mB_{\rm th}(M_1 \to M_2 + \inv) = \sum_{i,j=e,\mu,\tau}\mB(M_1 \to M_2 \nu_i \bar\nu_j).
\end{align}
The NP contributions to these processes can be generally described in the framework of SMEFT. We refer to refs.~\cite{Calibbi:2015kma,Allwicher:2023xba,Bause:2023mfe,Chen:2024jlj} for the recent relevant studies. In the following, we also focus on the SMEFT but with the MFV hypothesis.

In the SMEFT, as mentioned in section~\ref{sec:b2snunu}, one interesting implication of the MFV hypothesis is that the $d_i \to d_j \nu \bar \nu$ transitions are all governed by only one effective operator at low energy. As a consequence, the ratio of the branching fractions of any two $b \to s \nu \bar\nu$ decays should be independent of the NP contribution. Considering as an example the $B^+ \to K^+ \nu \bar\nu$ and $B^0 \to K^{*0} \nu \bar\nu$ decays, we obtain
\begin{align}\label{eq:MFV ratio:Kstar}
  \frac{\mB(B^+ \to K^+ \nu \bar\nu)}{\mB(B^0 \to K^{*0} \nu \bar\nu)} =  \frac{\mB(B^+ \to K^+ \nu \bar\nu)_\SM}{\mB(B^0 \to K^{*0} \nu \bar\nu)_\SM} = 0.46 \pm 0.07,
\end{align}
where summation over the neutrino flavours has been taken into account. In this ratio, all the relevant SMEFT operators, i.e. $\mQ_{lq}^{(1)}$ and $\mQ_{lq}^{(3)}$, have been included simultaneously.\footnote{The ratio still holds after including the operators $\mQ_{Hq}^{(1,3)}$, although they are not considered in our analysis as mentioned in section~\ref{sec:SMEFT}. 
} When deriving this ratio, it is noted that only the MFV hypothesis in the quark sector is required to eliminate the right-handed operator $\mO_R$. Therefore, the ratio still holds even by assuming generic couplings in the lepton sector, i.e., without the leptonic MFV. In addition, as both the CKM and Wilson coefficients involved are cancelled in this ratio, the theoretical uncertainty is reduced compared to that of the individual branching ratios. If future experimental measurements find any deviation from the theoretical value, it could imply new sources of flavour violation beyond the Yukawa couplings. By using the above ratio, the Belle $90\%$ CL bound $\mB(B^0 \to K^{*0} \nu \bar\nu) < 18 \times 10^{-6}$ would imply that
\begin{align}\label{eq:MFV prediction: K}
\mB(B^+ \to K^+ \nu \bar\nu)_\MFV < 10.5 \times 10^{-6},
\end{align}
which is obviously lower that the $90\%$ CL lower bound of the Belle II measurement, $\mB(B^+ \to K^+ \nu \bar\nu)>12.5 \times 10^{-6}$. Or inversely, the Belle II measurement $\mB(B^+ \to K^+ \nu \bar\nu)=(23_{-6}^{+7})\times 10^{-6}$ would indicate that
\begin{align}\label{eq:MFV prediction: Kstar}
\mB(B^0 \to K^{*0} \nu \bar\nu)_\MFV= \big( 50_{-16}^{+17} \big) \times 10^{-6},
\end{align}
which is incompatible with the Belle $90\%$ CL bound $\mB(B^0 \to K^{*0} \nu \bar\nu) < 18 \times 10^{-6}$. As a conclusion,  the recent Belle II measurement of the $B^+ \to K^+ \nu \bar\nu$ decay cannot be explained in the SMEFT with MFV hypothesis. However, the MFV prediction of $\mB(B^0 \to K^{*0} \nu \bar\nu)$ is well within the reach of Belle II at $5\ab^{-1}$ (cf. table~\ref{table:observables}), which can be used as a further test of the MFV hypothesis considered here.

Similarly, the ratio of the branching fractions of a $b \to s \nu \bar\nu$ and a $b \to d \nu \bar\nu$ decay is also independent of the NP contribution in the SMEFT with MFV hypothesis. As an illustration, let us consider the $B^+ \to K^+ \nu \bar\nu$ and $B^+ \to \pi^+ \nu \bar\nu$ decays, and obtain
\begin{align}\label{eq:MFV ratio:pi}
  \frac{\mB(B^+ \to K^+ \nu \bar\nu)}{\mB(B^+ \to \pi^+ \nu \bar\nu)} =   \frac{\mB(B^+ \to K^+ \nu \bar\nu)_\SM}{\mB(B^+ \to \pi^+ \nu \bar\nu)_\SM} = 29.7 \pm 5.6,
\end{align}
where summation over the neutrino flavours has been included. Taking as input the Belle II measurement in eq.~(\ref{eq:exp:Belle II}), this ratio implies that
\begin{align}
  \mB(B^+ \to \pi^+ \nu \bar\nu)_\MFV = \big( 7.8_{-2.6}^{+2.8} \big) \times 10^{-7},
\end{align}
which shows a $2.3\sigma$ deviation from the SM prediction $\mB(B^+ \to \pi^+ \nu \bar\nu)_\SM = (1.40 \pm 0.18)\times 10^{-7}$, but is still compatible with the current $90\%$ CL experimental bound $\mB(B^+ \to \pi^+ \nu \bar\nu) < 140 \times 10^{-7}$. Of course, similar ratios can also be constructed for other processes to test the framework used here. 

\begin{figure}[ht]
  \centering
  \includegraphics[width=0.45\linewidth]{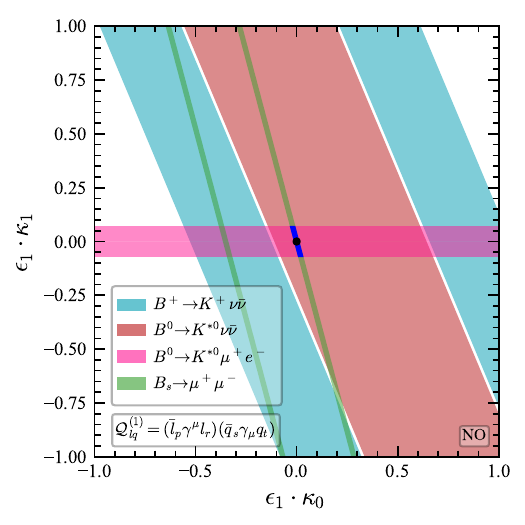}\quad
  \includegraphics[width=0.45\linewidth]{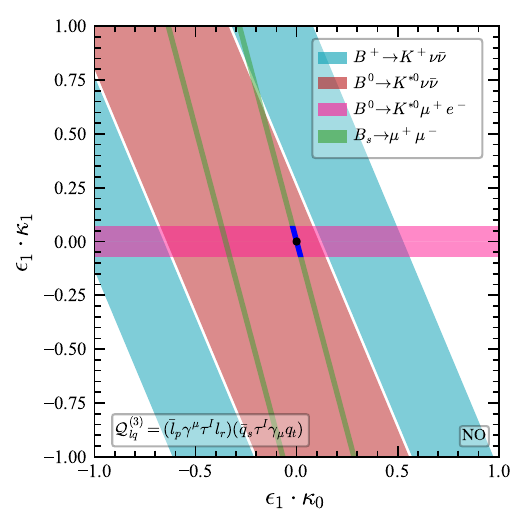}
  \\[-0.20cm]
  \includegraphics[width=0.45\linewidth]{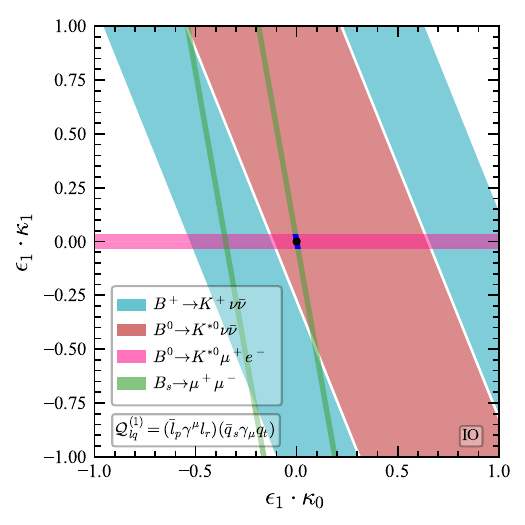}\quad
  \includegraphics[width=0.45\linewidth]{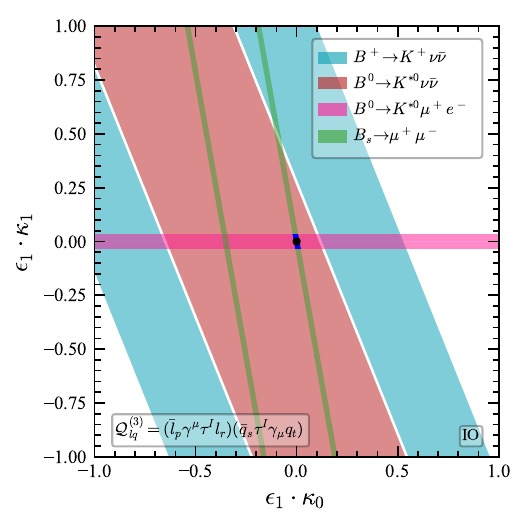}
  \caption{Constraints on the MFV parameter products $(\epsilon_1^{(i)}\kappa_0^{(i)}, \epsilon_1^{(i)}\kappa_1^{(i)})$ for $\mQ_i= \mQ_{lq}^{(1)}$ (left) and $\mQ_i = \mQ_{lq}^{(3)}$ (right) in the cases of NO (upper) and IO (lower) of neutrino masses. The allowed regions by the measurement of $\mB(B^+ \to K^+ \nu \bar\nu)$ are at $2\sigma$ level, while the regions by $\mB(B^0 \to K^{*0} \nu \bar\nu)$, $\mB(B^0 \to K^{*0} \mu^+ e^-)$ and $\mB(B_s \to \mu^+ \mu^-)$ at $90\%$ CL. The allowed regions by all the last three observables are shown in blue. The black points $(0,0)$ denote the SM case.}
  \label{fig:SMEFT:MFV}
\end{figure}

In the MFV framework, all the quark FCNC or the LFV processes are related with each other. Explicitly, for the operators $\mQ_{lq}^{(i)}$ ($i=1,3$), all the quark FCNC transitions are governed by the two parameter products $\epsilon_1^{(i)} \kappa_0^{(i)}$ and $\epsilon_1^{(i)} \kappa_1^{(i)}$. In order to constrain them,  we consider several quark (and lepton) FCNC processes, including the $B \to K^{(*)} \nu \bar \nu$, $B \to K^{(*)} e \mu$, $B \to \pi e \mu$, $B_s \to e \mu$, $B_s \to \mu^+ \mu^-$ as well as $K \to \pi \nu \bar\nu$ decays. Besides the input parameters listed in table~\ref{tab:input}, the lepton number breaking scale $\Lambda_{\rm LN}=10^{14}\GeV$ and the lightest neutrino mass $m_{\nu_1}=0.2\,\text{eV}$ are used in the leptonic MFV. Numerically, we find that the dominant constraints come from the $B^+ \to K^+ \nu \bar\nu$, $B^0 \to K^{*0} \nu \bar\nu$, $B^0 \to K^{*0} e^- \mu^+ $ and $B_s \to \mu^+ \mu^-$ decays. The allowed parameter regions of $\epsilon_1^{(i)} \kappa_0^{(i)}$ and $\epsilon_1^{(i)} \kappa_1^{(i)}$ are shown in figure~\ref{fig:SMEFT:MFV}. We can see that there is a small gap between the allowed regions of the $B^+ \to K^+ \nu \bar\nu$ and $B^0 \to K^{*0} \nu \bar\nu$ decays. This reflects the discrepancy between the MFV prediction in eq.~(\ref{eq:MFV prediction: Kstar}) and the current Belle bound on $\mB(B^0 \to K^{*0} \nu \bar\nu)$.

Since the uncertainty of the recent Belle II measurement is still quite large, we use instead all the processes except the $B^+ \to K^+ \nu \bar\nu$ decay to derive the allowed parameter space. As shown in figure~\ref{fig:SMEFT:MFV}, the upper bounds on the magnitudes of the products $\epsilon_1^{(i)} \kappa_0^{(i)}$ and $\epsilon_1^{(i)} \kappa_1^{(i)}$ for the two operators $\mQ_{lq}^{(1,3)}$ are all around $0.05$ in the cases of NO and IO of neutrino masses, which are much smaller than the parameter regions required to explain the $B^+ \to K^+ \nu \bar\nu$ excess. Using the allowed parameter regions for $\mQ_{lq}^{(i)}$, we can derive the following numerical predictions:
\begin{align}\label{eq:MFV prediction: K:2}
 3.4 \, (3.5) \times 10^{-6}< \mB(B^+ \to K^+ \nu \bar\nu)_\MFV &< 5.0 \, (4.9) \times 10^{-6},
  \nonumber\\
 7.4 \, (7.5) \times 10^{-6 } < \mB(B^0 \to K^{*0} \nu \bar\nu)_\MFV &< 10.8 \, (10.6) \times 10^{-6},
  \nonumber\\
 1.1 \, (1.2) \times 10^{-7} < \, \mB(B^+ \to \pi^+ \nu \bar\nu)_\MFV \, & < 1.7 \, (1.7) \times 10^{-7},
\end{align}
which correspond the NO (IO) of neutrino masses. The numerical results show less than $0.5\%$ difference between $\mQ_{lq}^{(1)}$ and $\mQ_{lq}^{(3)}$, implying that both of these two operators coincide with each other under the above bounds. From these bounds, we can see that the NP contributions to the $B\to K \nu \bar\nu$ and $B \to \pi \nu \bar\nu$ decays should be less than $\sim 20\%$ of their SM predictions. It is noted that the leptonic MFV has been used to derive the allowed parameter space and the above bounds. Therefore, the bound on $\mB(B^+ \to K^+ \nu \bar\nu)$ is more stringent than the one in eq.~(\ref{eq:MFV prediction: K}) derived without the leptonic MFV.

\subsection[$M_1 \to M_2  + \text{DM}$]{$\boldsymbol{M_1 \to M_2  + \text{DM}}$}
\label{sec:num:bsDM}

The large excess of $\mB(B^+ \to K^+ \nu \bar\nu)$ reported by Belle II can also be explained by light NP particles that contribute to the signal as $B^+ \to K^+ + \text{missing energy}$. In this case, the theoretical predictions of the $M_1 \to M_2 + \inv$ decay can be written as
\begin{align}
\mB_{\rm th}(M_1 \to M_2 + \text{inv})  = \mB_{\rm SM}(M_1 \to M_2 + \nu \bar\nu) + \mB_{\rm NP}(M_1 \to M_2 + \text{DM}).
\end{align}
As discussed in section~\ref{sec:theory:DM}, the effects of light NP particles in the second term can be generally described in the EFT framework. In the following, we investigate this explanation in the DSMEFT framework, in which the Wilson coefficients are taken to be the most general form or satisfy the MFV hypothesis.

\subsubsection{General flavour structure}

In the DSMEFT with general flavour structure, only the $b \to s + \inv$ processes are directly related to the $B^+ \to K^+ + \inv$ decays, which include $B^0 \to K^0 + \inv$, $B^{0,+} \to K^{* \, 0,+} + \inv$, $B_s \to \phi + \inv$ and $B_s \to \inv$ decays. In the following, we investigate constraints on the DSMEFT operators from the experimental data on these decays as summarized in table~\ref{table:observables}, and discuss future prospects at Belle II, CEPC and FCC-ee. As discussed in section~\ref{sec:DSMEFT}, seven DSMEFT operators are not hermitian. In order to compare with the MFV analysis in the next subsection, contributions from the hermitian conjugation of these operators are neglected in the numerical analysis in this subsection, i.e., $\mC_{ji}=0$ is taken for $d_i \to d_j$ processes.\footnote{Let us consider $\mQ_{d\phi^2}$ as an example. For the $b \to s \phi \phi$ process, $[\mC_{d\phi^2}]_{23} \neq 0$ and $[\mC_{d\phi^2}]_{32}=0$ are taken in the general DSMEFT analysis. In the MFV hypothesis, the coupling matrix eq.~(\ref{eq:MFV:num:2}) implies that $[\mC_{d\phi^2}]_{23} \gg [\mC_{d\phi^2}]_{32}$, which results in the approximation $[\mC_{d\phi^2}]_{32} \approx 0$. \label{note1}} For example, contributions to the $b \to s \phi \phi$ processes from the operator $\mQ_{d\phi^2}$ are calculated by taking $[\mC_{d\phi^2}]_{23}\neq 0$ and $[\mC_{d\phi^2}]_{32}=0$. Two exceptions are the operators $\mQ_{DqX^2}$ and $\mQ_{DdX^2}$. Consider the former as an example, we take $[\mC_{DqX^2}]_{ji}=[\mC_{DqX^2}]_{ij}^*$ for the $d_i \to d_j$ processes to compare with the MFV analysis, where its MFV coupling matrix in eq.~(\ref{eq:MFV:num:1}) also implies a hermitian $\mC_{DqX^2}$. In this case, the hadronic matrix elements of the weak currents can be parameterized in terms of the transition form factors of the vector current as discussed in appendix~\ref{sec:form factor}.

\begin{figure}[ht]
  \centering
  \includegraphics[width=0.43\linewidth]{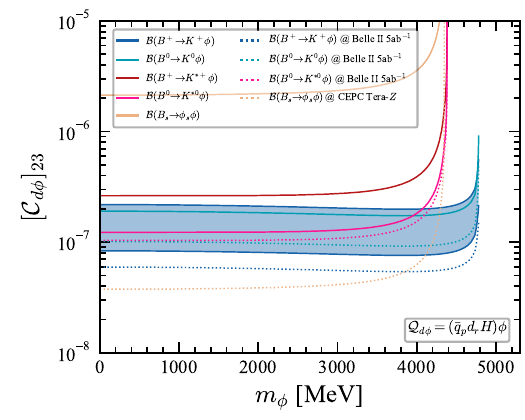}
  \qquad
  \includegraphics[width=0.43\linewidth]{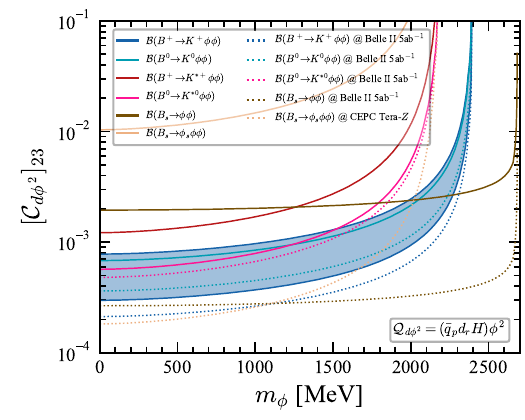}
  \\[0.20em]
  \includegraphics[width=0.43\linewidth]{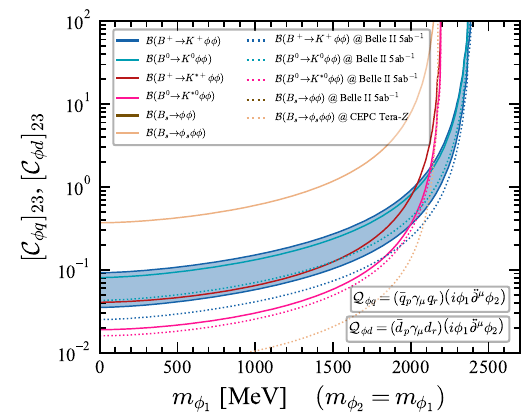}
  \qquad
  \includegraphics[width=0.43\linewidth]{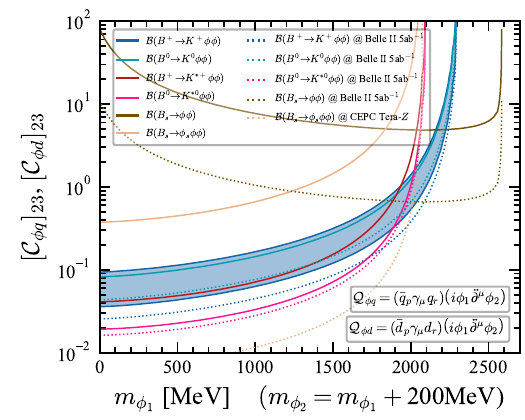}
  \\[0.20em]
  \includegraphics[width=0.43\linewidth]{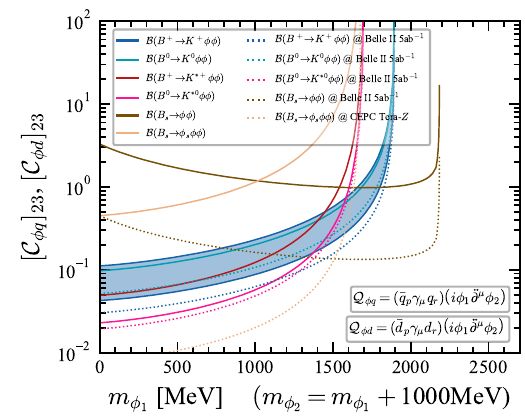}
  \qquad
  \includegraphics[width=0.43\linewidth]{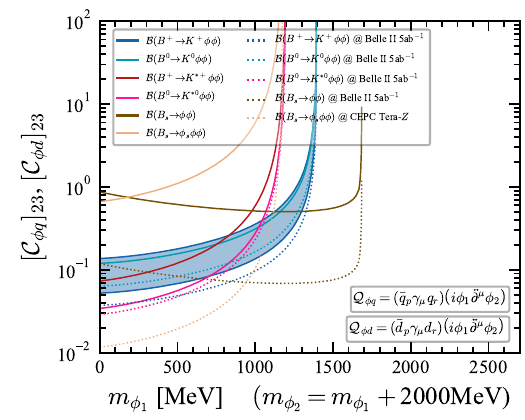}
  \caption{Constraints on the Wilson coefficients of the operators involving scalar DM as a function of the scalar mass $m_\phi$. The blue band denotes the $2\sigma$ allowed region by the recent Belle II measurement of the $B^+ \to K^+ + \inv$ decay. The upper bounds from other measurements are shown by the solid lines, while the expected sensitivities at Belle II and CEPC by the dotted lines. The sensitivities at FCC-ee are very similar to that at CEPC and hence are not shown. For the operators $\mQ_{\phi q}$ and $\mQ_{\phi d}$, we have considered four different cases for the DM masses, as indicated in the last four plots.}
  \label{fig:S}
\end{figure}

\begin{figure}[t]
  \centering
  \includegraphics[width=0.43\linewidth]{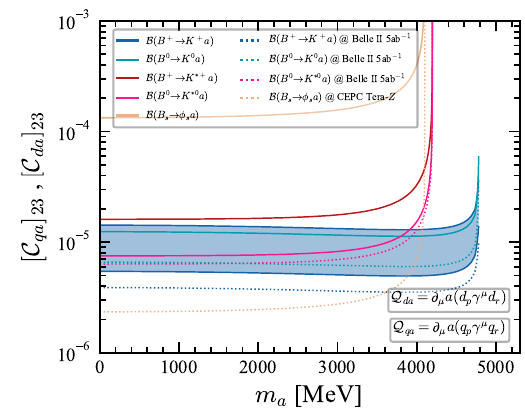}
  \caption{Same as in figure \ref{fig:S}, but for the operators involving ALP.}
  \label{fig:ALP}
\end{figure}

\begin{figure}[t]
  \centering
  \includegraphics[width=0.43\linewidth]{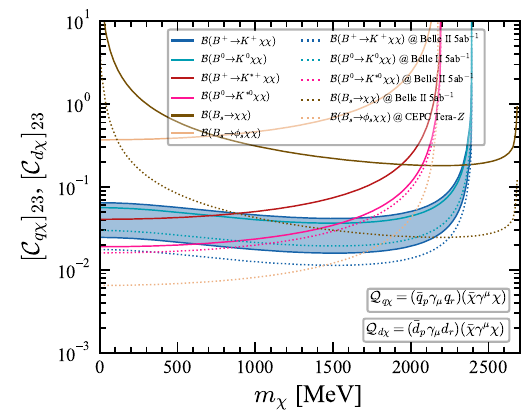}
  \caption{Same as in figure \ref{fig:S}, but for the operators involving fermionic DM.}
  \label{fig:F}
\end{figure}

For the scalar DM, there are three operators ($\mQ_{d\phi^2}$, $\mQ_{\phi q}$ and $\mQ_{\phi d}$, cf. eq.~(\ref{eq:DSMEFT:S})) contributing to the three-body $d_{i} \to d_{j} \phi \phi$ and one operator ($\mQ_{d\phi}$, cf. eq.~(\ref{eq:DSMEFT:S})) to the two-body $d_{i} \to d_{j} \phi$ decay. Allowed regions of the Wilson coefficients for these four operators are shown in figure~\ref{fig:S}. From this figure, we make the following observations:
\begin{itemize}
\item The $B^0 \to K^{*0} + \inv$ decay provides the strongest constraints on the parameter regions allowed by the Belle II measurement of the $B^+ \to K^+ + \inv$ decay. Especially, for the operators $\mQ_{\phi q}$ and $\mQ_{\phi d}$, the regions of $m_{\phi_2} \lesssim 2.0 \GeV$ ($m_{\phi_1} < m_{\phi_2}$ is assumed without loss of generality) are excluded by the decay. For the operator $\mQ_{d \phi}$, on the other hand, large part of the parameter region with $m_\phi \lesssim 3.8 \GeV$ is excluded.

\item Another stronger constraint arises from the $B_s \to \inv$ decay, which benefits from the large phase space. As a result, it excludes the parameter region for large DM mass, e.g., the region with $m_\phi \gtrsim 2.3\GeV$ is excluded for the operator $\mQ_{d \phi^2}$. For the operators $\mQ_{\phi q}$ and $\mQ_{\phi d}$, however, the $B_s \to \phi_1 \phi_2$ amplitude vanishes in the case of $m_{\phi_1}=m_{\phi_2}$ due to the derivative term $\phi_1 \overleftrightarrow{\partial^\mu} \phi_2$ in the operators. Therefore, only the parameter regions with $m_{\phi_1} \neq m_{\phi_2}$ receive constraint from the $B_s \to \inv$ decay.

\item Compared to the $B^0 \to K^{*0} + \inv$ and $B_s \to \inv$ decays, the $B^0 \to K^0 + \inv$, $B^+ \to K^{*+} + \inv$ and $B_s \to \phi + \inv$ decays provide almost no further constraints. The upper bounds from these three decays are comparable to or weaker than the upper limits of the $2\sigma$ allowed region required to explain the $B^+ \to K^+ + \inv$ excess.

\item All the processes put very weak constraints on the parameter regions with $m_\phi \lesssim 2.3 \GeV$ for the operator $\mQ_{d\phi^2}$.
\end{itemize}
In addition, constraints on the operators involving ALP are shown in figure~\ref{fig:ALP}, which are very similar to the ones for the operator $\mQ_{d\phi}$ but with much larger allowed Wilson coefficients.

For the fermionic DM, there are only two relevant operators $\mQ_{q\chi}$ and $\mQ_{d\chi}$ (cf. eq.~(\ref{eq:DSMEFT:F})) in the DSMEFT. The experimental constraints are shown in figure~\ref{fig:F}. In the cases of $m_\chi \lesssim 0.6\GeV$ and $m_\chi \gtrsim 2.3\GeV$, the parameter spaces required to account for the Belle II measurement are already excluded by the $B^0 \to K^{*0} + \inv$ and $B_s \to \inv$ decays, respectively. In the range of $0.6 \lesssim m_\chi \lesssim 2.3\GeV$, however, large part of the parameter space survives all the constraints. This is quite different from the case of the SMEFT operator $\mQ_{lq}^{(1)}$ discussed in section~\ref{sec:num:bsnunu}, due to the massive fermionic DM fields in the operators $\mQ_{q\chi}$ and $\mQ_{d \chi}$. In the massless limit, their effects on the decay width are the same as that of the operator $\mQ_{lq}^{(1)}$.\footnote{Although the leptonic and DM currents in these operators have different chiralities, the total decay width does not depend on the chirality of the currents.}

\begin{figure}[t]
  \centering
  \includegraphics[width=0.43\linewidth]{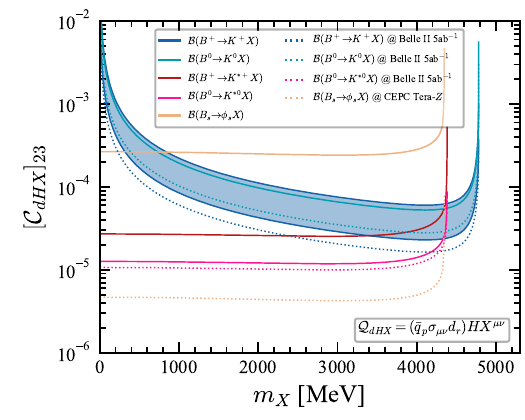}
  \qquad
  \includegraphics[width=0.43\linewidth]{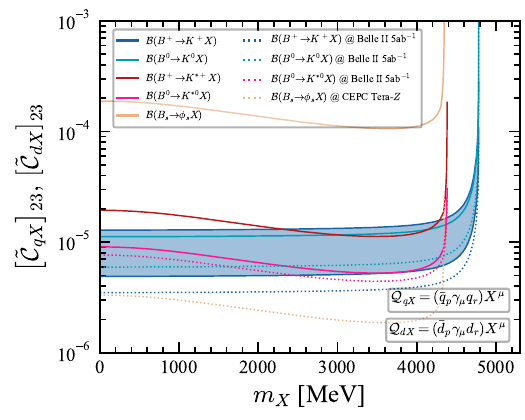}
  \\[0.20em]
  \includegraphics[width=0.43\linewidth]{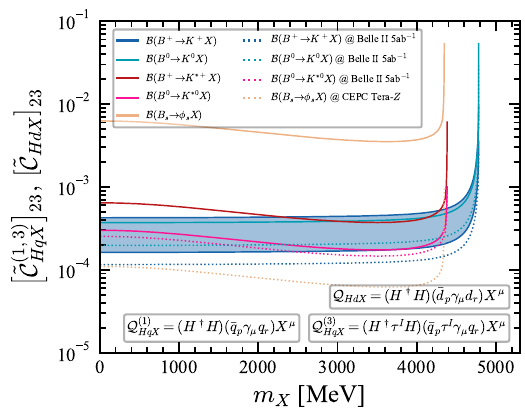}
  \caption{Same as in figure \ref{fig:S}, but for the operators involving one vector DM.}
  \label{fig:V}
\end{figure}

\begin{figure}
  \centering
  \includegraphics[width=0.43\linewidth]{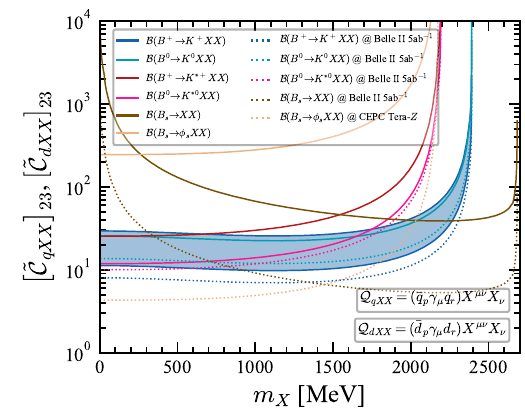}
  \qquad
  \includegraphics[width=0.43\linewidth]{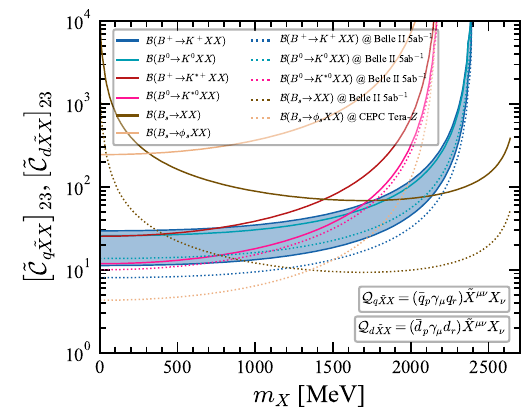}
  \\[0.0em]
  \includegraphics[width=0.43\linewidth]{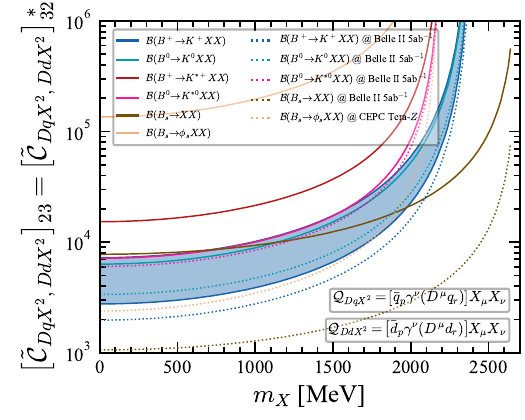}
  \qquad
  \includegraphics[width=0.43\linewidth]{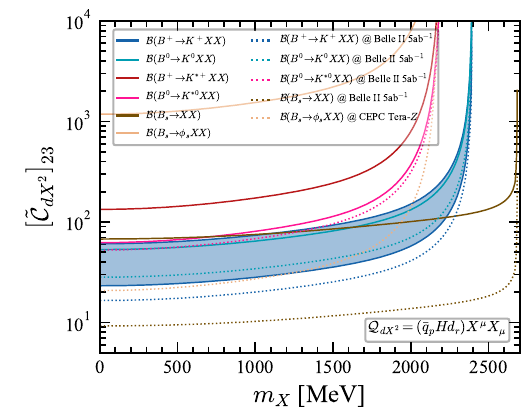}
  \\[0.0em]
  \includegraphics[width=0.43\linewidth]{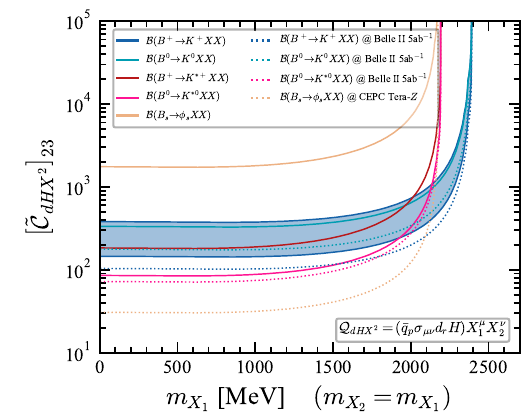}
  \qquad
  \includegraphics[width=0.43\linewidth]{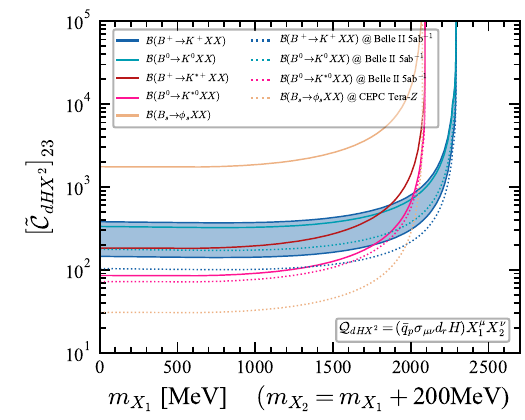}
  \\[0.0em]
  \includegraphics[width=0.43\linewidth]{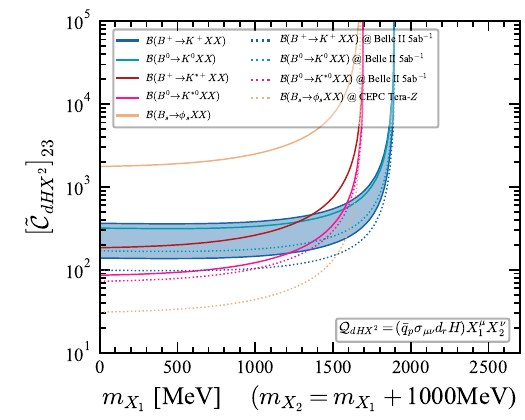}
  \qquad
  \includegraphics[width=0.43\linewidth]{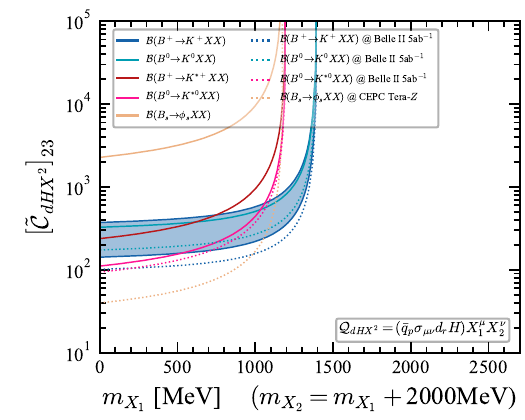}
  \caption{Same as in figure \ref{fig:S}, but for the operators involving two vector DM fields. For the operator $\mQ_{dHX^2}$, we have considered four different cases for the masses of the two DM particles, as indicated in the last four plots.}
  \label{fig:V:2}
\end{figure}

For the vector DM, there are 8 operators ($\mQ_{dX^2}$, $\mQ_{qXX}$, $\mQ_{dXX}$, $\mQ_{q \tilde X X}$, $\mQ_{d \tilde X X}$, $\mQ_{DqX^2}$, $\mQ_{DdX^2}$ and $\mQ_{dHX^2}$, cf. eq.~(\ref{eq:DSMEFT:V:2})) contributing to the three-body $d_{i} \to d_{j} X X$ and 6 operators ($\mQ_{dHX}$, $\mQ_{dX}$, $\mQ_{qX}$, $\mQ_{HdX}$ and $\mQ_{HqX}^{(1,3)}$, cf. eqs.~(\ref{eq:DSMEFT:V:1}) and (\ref{eq:DSMEFT:V:2})) to the two-body $d_{i} \to d_{j} X$ decay. For the Wilson coefficient of each operator, the parameter region required to account for the Belle II measurement of the $B^+ \to K^+ + \inv$ decay as well as the experimental bounds from other processes are shown in figures~\ref{fig:V} and \ref{fig:V:2}. The following observations are made:
\begin{itemize}
\item For the operators involving one DM field, the high-mass region of $4.3 \lesssim m_X\lesssim 4.6\GeV$ does not suffer from any constraint, while large part of the parameter region with $m_X \lesssim 4.3\GeV$ is excluded by the $B^0 \to K^{*0} + \inv $ decay. Specifically, for the operator $\mQ_{dHX}$, the unique one built from the field strength tensor $X_{\mu\nu}$, the parameter region with $m_X\lesssim 4.3\GeV$ is excluded; for the other operators, parts of the parameter region still survive for $m_X \lesssim 2.5\GeV$ or $4.3 \lesssim m_X < 4.6 \GeV$.

\item For the operators involving two DM fields, except for $\mQ_{dHX^2}$, the parameter spaces with $m_X \gtrsim 2\GeV$ are all excluded by the $B_s \to \inv$ decay. For $\mQ_{DqX^2}$, $\mQ_{DdX^2}$ and $\mQ_{dX^2}$, the $B_s \to \inv$ decay also provides almost the strongest constraints in the whole DM mass region, although the constraints in the region of $m_X \lesssim 1.7\GeV$ are quite weak. For the other operators, the parameter regions with $m_X \lesssim 1.7\GeV$ are highly constrained or largely excluded by the $B^0 \to K^{*0} + \inv$ decay.

\item Constraints on the operator $\mQ_{dHX^2}$ are quite different from the other ones. As the tensor hadronic matrix element $\langle 0|\bar{b}\sigma_{\mu\nu}s|B_s\rangle$ vanishes identically, this operator does not contribute to the $B_s \to \inv$ decay. Therefore, the high-mass regions do not suffer from any constraint. However, most of the low-mass regions are excluded by the $B^0 \to K^{*0} + \inv$ decay, e.g., the one with $m_{X_1}\lesssim 1.8\GeV$ is excluded in the case of $m_{X_2}-m_{X_1}=200\MeV$.
\end{itemize}

All the data on $b\to s$ decays discussed above will be significantly improved in the future~\cite{Belle-II:2022cgf,Li:2022tov,Belle-II:2018jsg,Amhis:2023mpj}. Considering the expected sensitivities at Belle II ($5\ab^{-1}$), CEPC and FCC-ee (Tera-$Z$ phase) listed in table~\ref{table:observables}, we also show in figures~\ref{fig:S}-\ref{fig:V:2} the projected $90\%$ CL bounds on the DSMEFT operators. It is observed that
\begin{itemize}
\item At the Belle II with $5\ab^{-1}$, the relative uncertainty of the measured $\mB(B^+ \to K^+ \nu \bar\nu)$ is expected to be $19\%$. Using the SM prediction to estimate the future experimental central value, the projected bound covers all the parameter space explaining the recent Belle II excess. If the central value of the current Belle II data remains unchanged in the future, the future experimental measurement of $\mB(B^+ \to K^+ \nu \bar\nu)$ is expected to be $(23 \pm 4) \times 10^{-6}$, which would deviate from the current SM prediction by about $4\sigma$.

\item For the $B^0 \to K^0 +\inv$ decay, the projected bound can cover most of the parameter regions required to explain the $B^+ \to K^+ + \inv$ excess. Therefore, this decay can provide an independent probe of the proposed mechanism behind the Belle II excess. 

\item The $B_s \to \phi + \inv$ and $B_s \to \inv$ decays can strongly constrain the low- and high-mass regions of the parameter space used to explain the Belle II excess, respectively. Their combination can even exclude almost all the parameter regions. It is also noted that the operators $\mQ_{d\phi^2}$ and $\mQ_{dX^2}$ affect the $B_s \to \inv$ decay without helicity suppression, which makes the projected bound of the $B_s \to \inv$ decay cover the whole parameter space of these two operators.

\item All the operators involving one DM (such as $\mQ_{d\phi}$ and $\mQ_{da}$), as well as the operator $\mQ_{dHX^2}$, do not contribute to the $B_s \to \inv$ decay. Therefore, the high-mass regions do not suffer from any constraint, even at the Belle II with $5\ab^{-1}$ or at the CEPC and FCC-ee during the Tera-$Z$ phase.
\end{itemize}

\subsubsection{MFV}

\begin{figure}[t]
  \centering
  \includegraphics[width=0.495\linewidth]{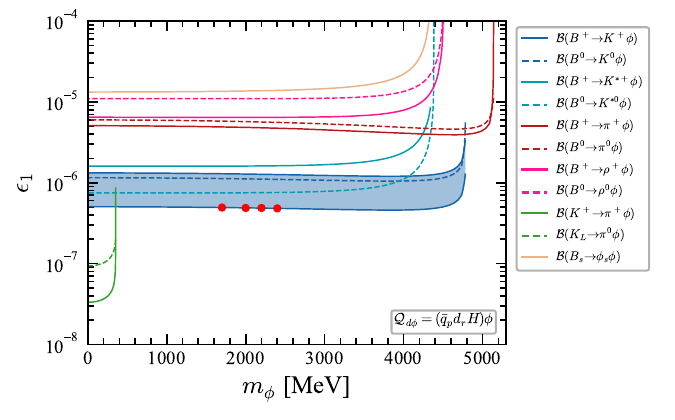}
  \includegraphics[width=0.495\linewidth]{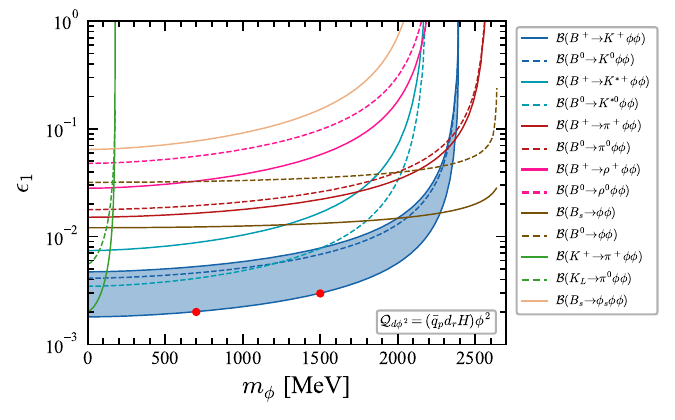}
  \\[0.20em]
  \includegraphics[width=0.495\linewidth]{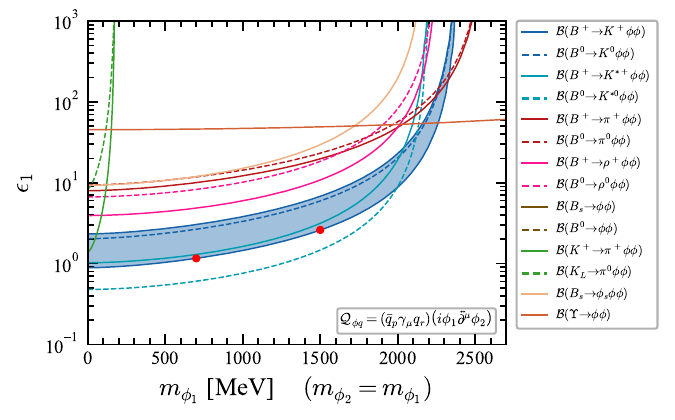}
  \includegraphics[width=0.495\linewidth]{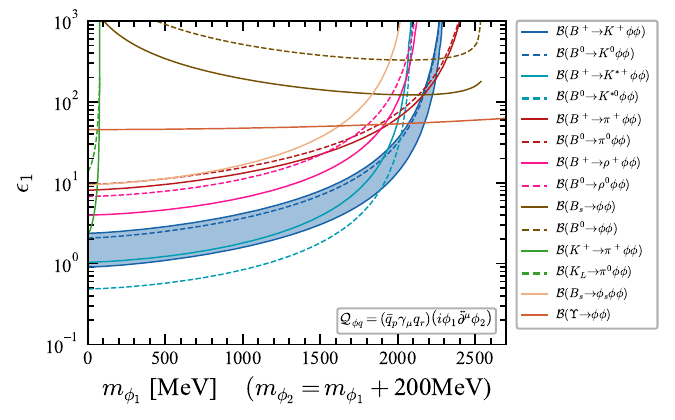}
  \\[0.20em]
  \includegraphics[width=0.495\linewidth]{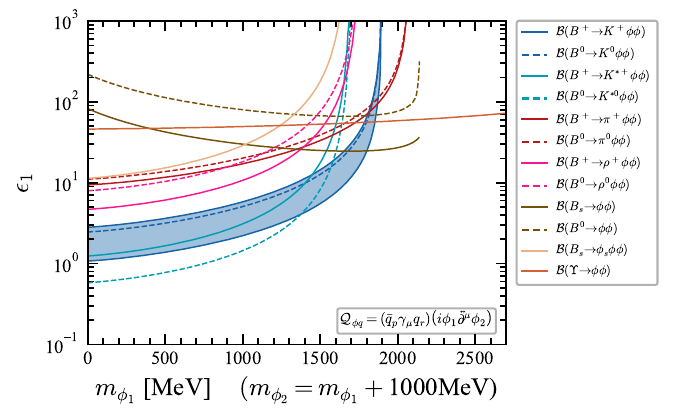}
  \includegraphics[width=0.495\linewidth]{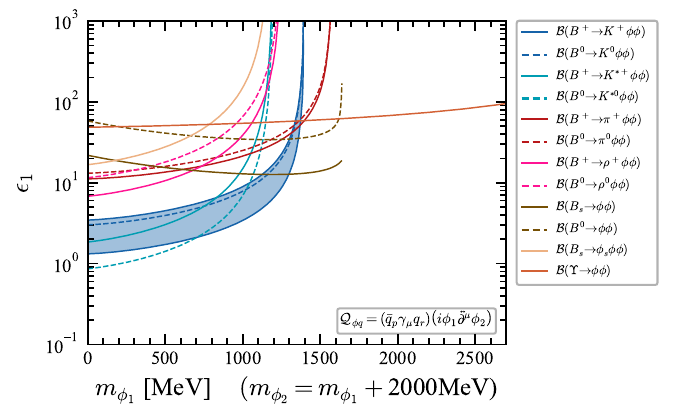}
  \caption{Constraints on the MFV parameter $\epsilon_1$ for the operators involving scalar DM as a function of the scalar mass $m_\phi$. The blue band denotes the $2\sigma$ allowed region by the recent Belle II measurement of the $B^+ \to K^+ + \inv$ decay, while the upper bounds obtained from all the other processes are shown by the solid and dashed lines. The red points denote the benchmark points used to investigate the $q^2$ distributions in figure~\ref{fig:BR}; see text for more details.}
  \label{fig:S:MFV}
\end{figure}

As demonstrated in eq.~(\ref{eq:WC:MFV}), once the MFV hypothesis is assumed, 8 out of the 22 DSMEFT operators do not induce down-type FCNC interactions. Therefore, these operators cannot account for the recent Belle II measurement of the $B^+ \to K^+ + \inv $ decay. For the remaining 14 operators, their MFV couplings make the $b \to d$ and $s \to d$ processes also relevant. We consider the constraints from the experimental measurements summarized in table~\ref{table:observables}, which include the branching ratios of the $B \to K^{(*)}, \pi, \rho + \inv$, $K^+ \to \pi ^+ + \inv$, $K_L \to \pi^0 + \inv$, $B_s \to \phi + \inv$ and $B_{s,d} \to \inv$ decays. It is noted that, as discussed in the last subsection, the $b \to s $ transition in the MFV hypothesis approximately depends only on one Wilson coefficient. Numerically, this situation is similar to the treatment in the general DSMEFT discussed in the last section; see footnote~\footref{note1} for an example. Consequently, the relative strength of the constraints from the various $b \to s$ processes are the same between the general case and the MFV hypothesis. Therefore, in the following, we focus on whether the $s \to d$ and $b \to d$ processes can provide further constraint with respect to that from the $b \to s$ processes.

\begin{figure}[t]
  \centering
  \includegraphics[width=0.495\linewidth]{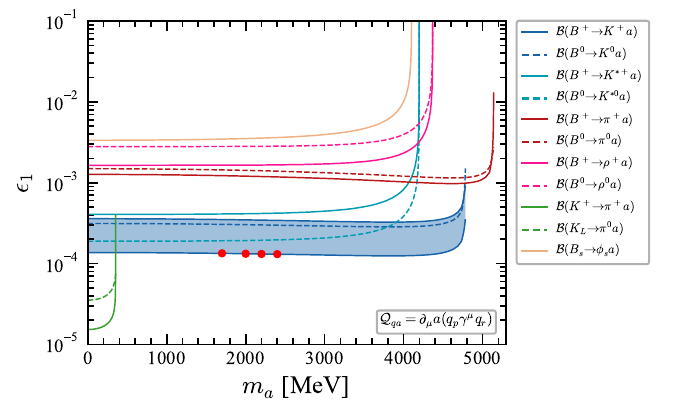}
  \caption{Same as in figure~\ref{fig:S:MFV}, but for the operator involving ALP.}
  \label{fig:ALP:MFV}
\end{figure}

\begin{figure}[t]
  \centering
  \includegraphics[width=0.495\linewidth]{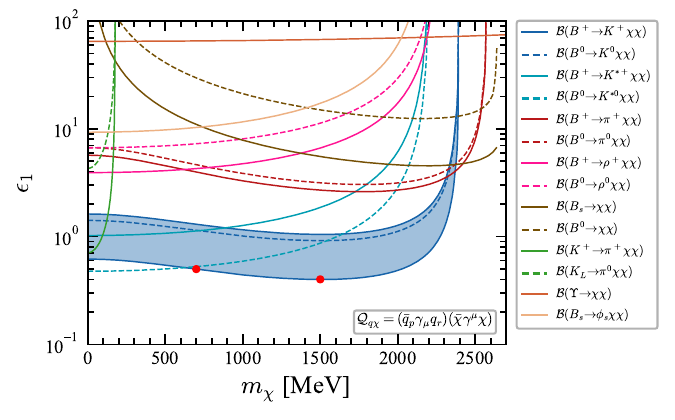}
  \caption{Same as in figure~\ref{fig:S:MFV}, but for the operator involving fermionic DM.}
  \label{fig:F:MFV}
\end{figure}

For the three operators with scalar DM (cf. eq.~(\ref{eq:DSMEFT:S})), the allowed regions of the MFV parameter $\epsilon_1^i$ are shown in figure~\ref{fig:S:MFV}. In the low-mass region of $m_\phi\lesssim 200\MeV$, the $K^+ \to \pi^+ + \inv$ decay puts the strongest constraint for the operators $\mQ_{d \phi}$ and $\mQ_{d\phi^2}$. For all the operators, the $B^+ \to \pi^+ + \inv$ or the $B^+ \to \rho^+ + \inv$ decay, depending on the DM mass, provides the strongest constraint among all the $b \to d $ processes. However, this decay provides no further constraint after considering the constraints from the $b \to s $ processes. The only exception is the constraint for the operator $\mQ_{\phi q}$ in the case of $m_{\phi_1}= m_{\phi_2} \equiv m_\phi$. In the region of $m_\phi \gtrsim 2.3\GeV$, the strongest constraint is provided by the $B^+ \to \pi^+ + \inv$ decay (as well as the $\Upsilon(1S) \to \inv$ decay; see the later discussion), because $m_{\phi_1} = m_{\phi_2}$ makes the NP contribution to the $B_s \to \phi_1 \phi_2$ decay vanishing. In addition, $\mQ_{qa}$ is the unique operator involving ALP in the MFV framework. The constraints on this operator are shown in figure~\ref{fig:ALP:MFV}, which are, as in the general case, very similar to the ones for $\mQ_{d\phi}$ but with a much larger allowed Wilson coefficient.

Among the operators with fermionic DM, there is only one relevant operator ($\mQ_{q\chi}$, cf. eq.~(\ref{eq:DSMEFT:F})) in the MFV hypothesis. The experimental constraints are shown in figure~\ref{fig:F:MFV}. We can see that, after considering the constraints from the $b \to s $ processes, all the $s \to d$ and $b \to d$ decays provide no further constraints, except in a narrow window near $m_\chi \approx 2.3\GeV$, where the $B^+ \to \pi^+ + \inv$ decay gives the strongest bound.

\begin{figure}[t]
  \centering
  \includegraphics[width=0.495\linewidth]{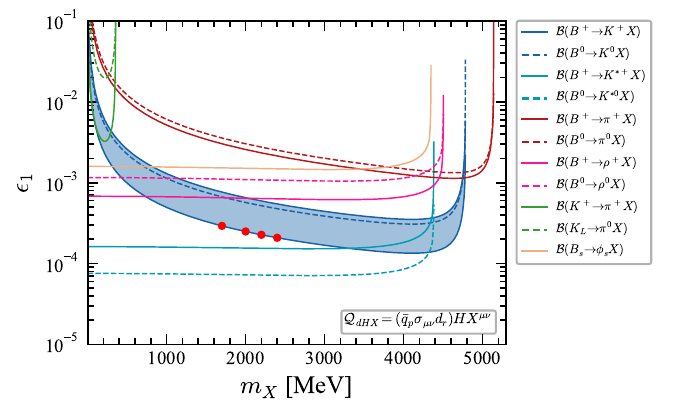}
  \includegraphics[width=0.495\linewidth]{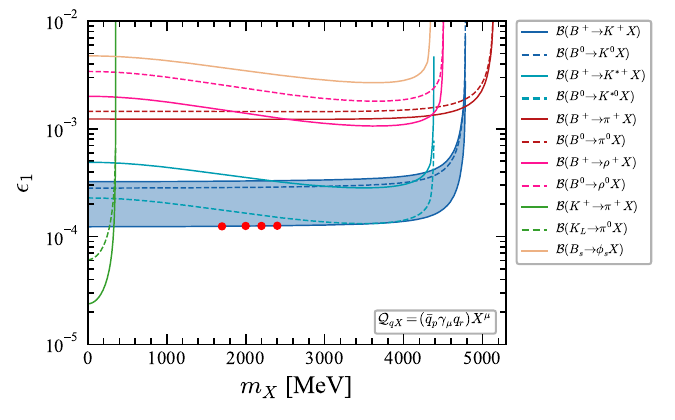}
  \\[0.20em]
  \includegraphics[width=0.495\linewidth]{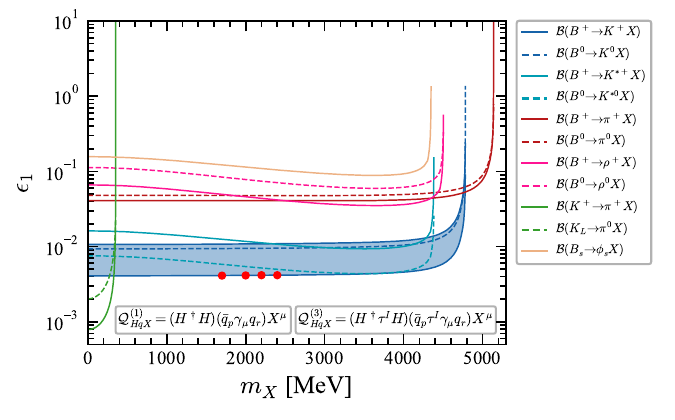}
  \caption{Same as in figure~\ref{fig:S:MFV}, but for the operators involving one vector DM.}
  \label{fig:V:MFV}
\end{figure}

\begin{figure}
  \centering
  \includegraphics[width=0.495\linewidth]{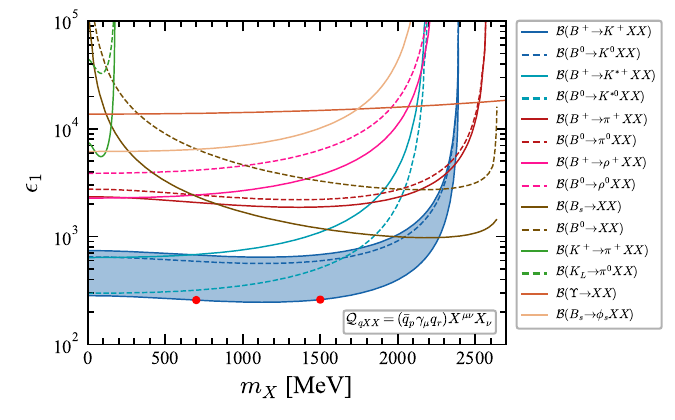}
  \includegraphics[width=0.495\linewidth]{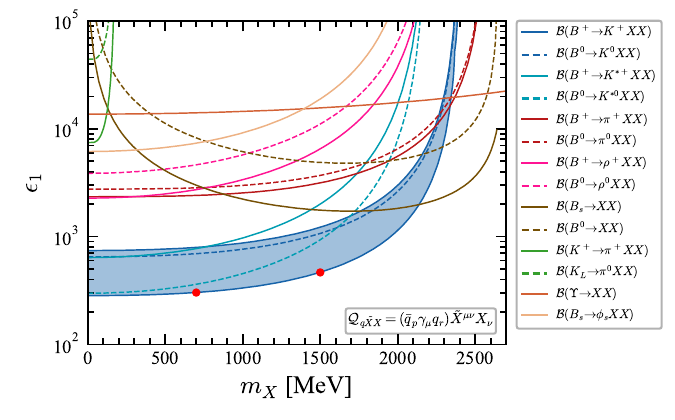}
  \\[0.2em]
  \includegraphics[width=0.495\linewidth]{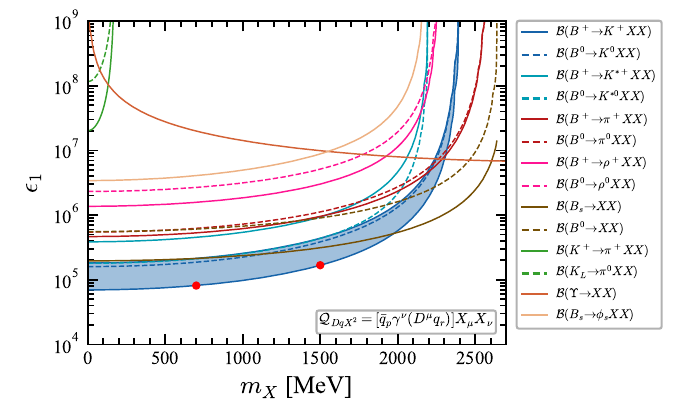}
  \includegraphics[width=0.495\linewidth]{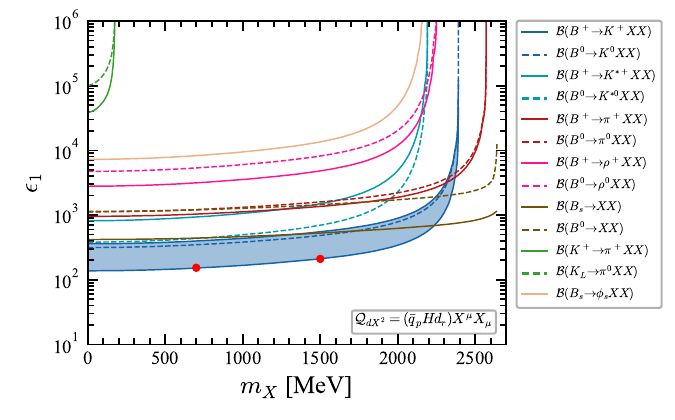}
  \\[0.2em]
  \includegraphics[width=0.495\linewidth]{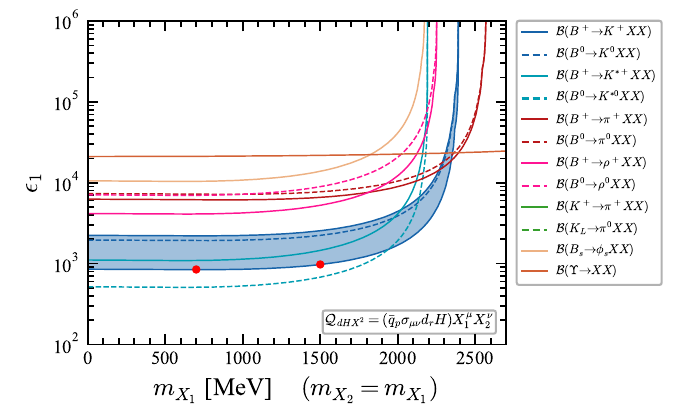}
  \includegraphics[width=0.495\linewidth]{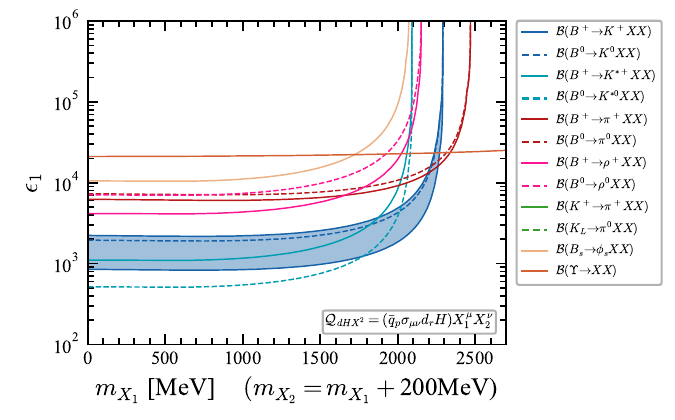}
  \\[0.2em]
  \includegraphics[width=0.495\linewidth]{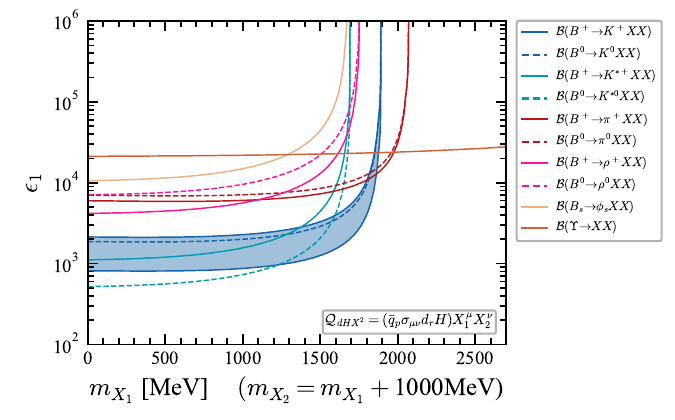}
  \includegraphics[width=0.495\linewidth]{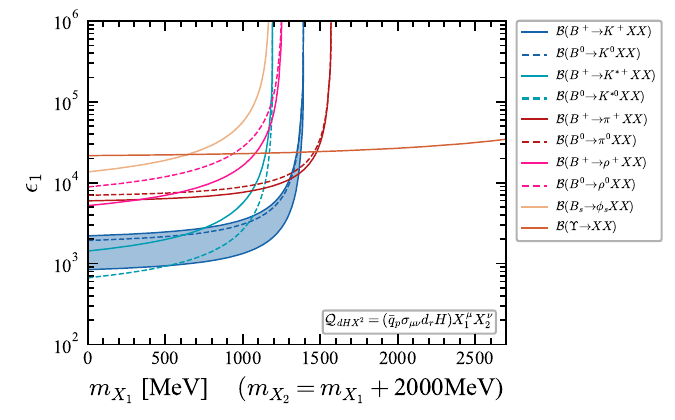}
  \caption{Same as in figure~\ref{fig:S:MFV}, but for the operators involving two vector DM fields. For the operator $\mQ_{dHX^2}$, we have considered four different cases for the DM masses, as indicated in the last four plots.}
  \label{fig:V:MFV:2}
\end{figure}

For the vector DM, in the MFV framework, there are five operators ($\mQ_{qXX}$, $\mQ_{q \tilde X X}$, $\mQ_{dX^2}$, $\mQ_{DqX^2}$ and $\mQ_{dHX^2}$, cf. eq.~(\ref{eq:DSMEFT:V:2})) contributing to the three-body $d_{i} \to d_{j} X X$ and four operators ($\mQ_{dHX}$, $\mQ_{qX}$ and $\mQ_{HqX}^{(1,3)}$, cf. eqs.~(\ref{eq:DSMEFT:V:1}) and (\ref{eq:DSMEFT:V:2})) to the two-body $d_{i} \to d_{j} X$ decay. The relevant experimental constraints are shown in figures~\ref{fig:V:MFV} and \ref{fig:V:MFV:2}. For the operators $\mQ_{qX}$ and $\mQ_{HqX}^{(1,3)}$, the $K^+ \to \pi^+ + \inv $ decay provides the strongest constraint for $m_X \lesssim 0.35\GeV$ and exclude the parameter region required to account for the Belle II excess. For the four operators involving one DM, the $B^+ \to \pi^+ + \inv$ decay excludes the mass window around $m_X \approx 4.8\GeV$. For the five operators involving two DM, neither the $s \to d$ nor the $b \to d$ processes can put further constraints after considering those from the $b \to s$ decays. One exception is the operator $\mQ_{dHX^2}$, where the decay $B^+ \to \pi^+ + \inv$ excludes the tail in the high-mass region.

In the MFV hypothesis, as indicated by eqs.~(\ref{eq:MFV:num:1}) and (\ref{eq:MFV:num:2}), the flavour-conserving couplings to the third generation are much larger than all the other FCNC couplings.\footnote{As can be seen from eq.~(\ref{eq:WC:MFV}), the flavour-conserving couplings can also be altered by the MFV parameter $\epsilon_0$. However, this parameter does not affect the FCNC processes, e.g., the $B^+ \to K^+ + \inv$ decay. For simplicity, we take $\epsilon_0 = 0 $ to derive the bounds on $\epsilon_1$ from the invisible decays of charmoniums and bottomoniums. For a non-zero $\epsilon_0$, it is straightforward to obtain the corresponding bounds on $\epsilon_1$ by using eq.~(\ref{eq:WC:MFV}).} Therefore, the $\Upsilon \to \DM + \DM$ decay could be highly enhanced. However, the current experimental bounds on the invisible $\Upsilon$ decays are still quite weak. For the $\Upsilon(1S) \to \inv$ decay, the $90\%$ CL upper limit on its branching ratio is $3.0 \times 10^{-4}$~\cite{BaBar:2009gco}, which is about one order of magnitude higher than the SM prediction $9.9\times 10^{-6}$~\cite{Chang:1997tq,Fernandez:2014eja,Bertuzzo:2017lwt}. Constraints from this process are also shown in figures~\ref{fig:S:MFV}-\ref{fig:V:MFV:2}. For the operator $\mQ_{\phi q}$ with $m_{\phi_2}-m_{\phi_1} \leq 200 \MeV$, we can see that the constraints from the $\Upsilon(1S) \to \inv$ decay are much stronger than from all the $b \to d$ decays in the high mass region of $m_{\phi_2} \gtrsim 2\GeV$ and, especially, exclude the regions of $m_{\phi_2} \gtrsim 2.3\GeV$ required to explain the Belle II excess. However, the constraints for all the other operators are much weaker than from the $B$-meson and kaon decays in the range of $m_\DM \lesssim m_B/2$. On the other hand, the constraints from invisible charmonium decays are totally negligible, because the NP contributions are suppressed by a factor of $\mO(\lambda_c^4)$ in the MFV hypothesis.

\begin{figure}[t]
  \centering
  \includegraphics[width=0.49\linewidth]{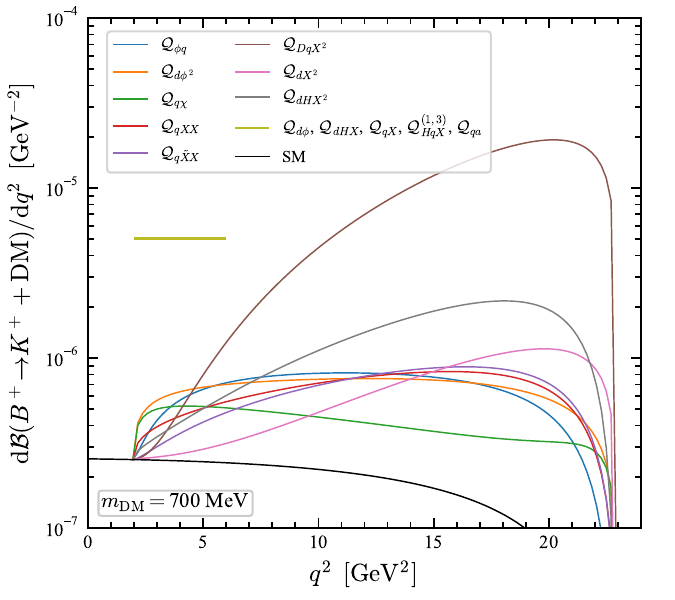}
  \includegraphics[width=0.49\linewidth]{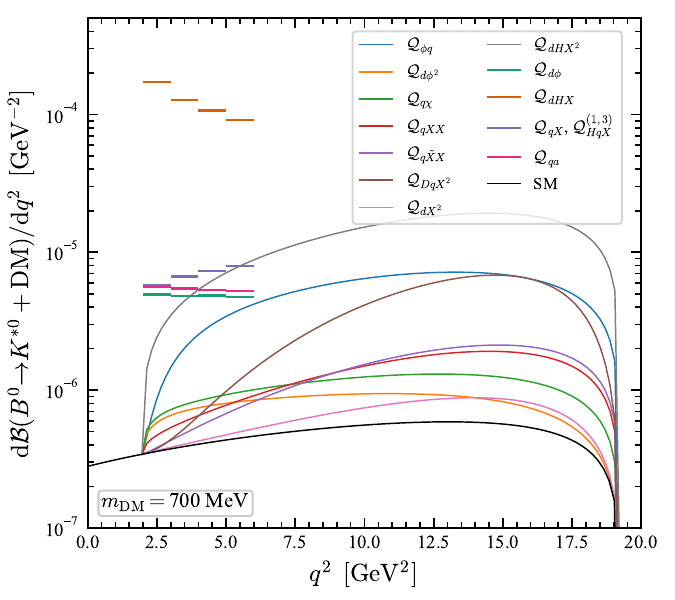}
  \\[0.20em]
  \includegraphics[width=0.49\linewidth]{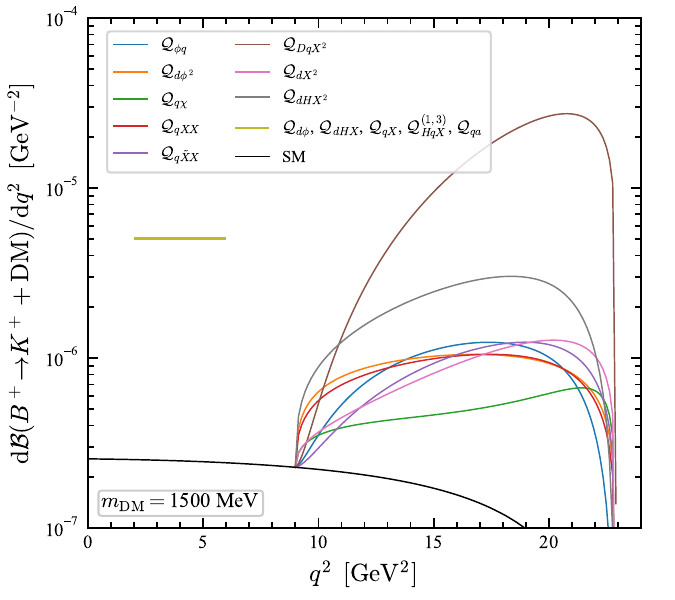}
  \includegraphics[width=0.49\linewidth]{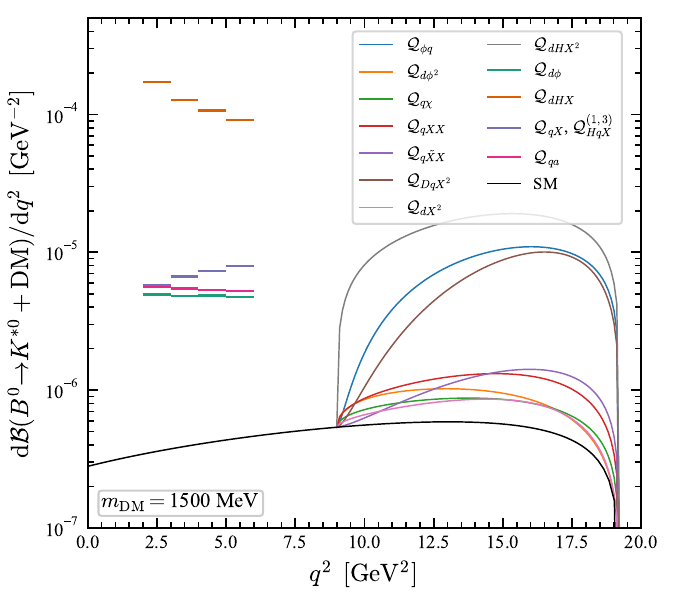}
  \caption{Differential branching ratios of the $B^+ \to K^+ + \inv$ and $B^0 \to K^{*0} + \inv$ decays both within the SM and in the DSMEFT with MFV hypothesis. For each effective operator, the DM mass is chosen either as $700\MeV$ or $1500\MeV$, and the MFV parameters are fixed at the corresponding benchmark points shown in figures~\ref{fig:S:MFV}--\ref{fig:V:MFV:2}. For the operators involving one DM field, $m_\DM=1.7$, $2.0$, $2.2$ and $2.4\GeV$ are chosen and the corresponding averaged branching ratios in the $[2,\,3]$, $[3,\, 4]$, $[4,\, 5]$ and $[5,\, 6]\GeV^2$ bins are shown, respectively.}
  \label{fig:BR}
\end{figure}

The $q^2$ distributions of the FCNC decays can also provide useful information about the NP effects. Considering the large uncertainty of the Belle II measurement, we take a conservative estimation of the NP contributions to the $q^2$ distributions: For each DSMEFT operator, the minimal value of the MFV parameter allowed by all the experimental constraints is chosen as the benchmark point, which is also shown in figures~\ref{fig:S:MFV}-\ref{fig:V:MFV:2}. Therefore, all the benchmark points actually correspond to the same branching ratio $\mB(B^+ \to K^+ + \inv)=4.8\times 10^{-6}$, which is the $2\sigma$ lower bound allowed by the Belle II measurement. Furthermore, we choose two typical values of the DM masses, $700\MeV$ and $1500\MeV$. By using these benchmark points, we show in figure~\ref{fig:BR} our predictions for the $q^2$ distributions of $B^+ \to K^+ + \inv$ and $B^0 \to K^{*0} + \inv$ decays.\footnote{The $q^2$ distributions of other $P\to P + \inv$ and $P \to V + \inv$ decays are similar to that of $B^+ \to K^+ + \inv$ and $B^0 \to K^{*0} + \inv$ respectively, and hence are not shown here anymore. In addition, for comparison, we show the $q^2$ distributions of these decays by considering the benchmark values of the Wilson coefficients of the operators $\mQ_{\phi q}$, $\mQ_{dHX}$ and $\mQ_{dHX^2}$, although they are already excluded by the $B^0 \to K^{*0} + \inv$ bound.} In the case of $m_\DM=700\MeV$, we can see that, for the operators $\mQ_{\phi q}$ and $\mQ_{d\phi^2}$, the distributions are close to each other in the $B^+ \to K^+ + \inv$, but distinguishable in the $B^0 \to K^{*0}+ \inv$ decay. However, for the operators $\mQ_{qXX}$ and $\mQ_{q\tilde{X}X}$, the distributions are quite close in both of these two decays. In the case of $m_\DM=1500\MeV$, the distributions are quite different, and hence distinguishable from each other for most operators. Therefore, all the operators are distinguishable from each other by combining the two decay spectra.

As can be seen from ref.~\cite{Belle-II:2023esi}, excesses of the $B^+ \to K^+ + \inv$ events mostly appear in the bins $2\leq q^2 \leq 7 \GeV^2$, although the uncertainties are still quite large. From the predictions shown in figure~\ref{fig:BR}, one can see that the $q^2$ distribution resulting from the operator $\mQ_{q\chi}$ with $m_\chi=700\MeV$ can closely match that of the Belle II excess. As another possibility, the two-body decay with a DM mass around $2\GeV$ could also contribute to the observed distribution. In this case, the DSMEFT operators involving one DM field become relevant, and there are 6 such operators in the MFV hypothesis, i.e., $\mQ_{d\phi}$, $\mQ_{qa}$, $\mQ_{dHX}$, $\mQ_{qX}$ and $\mQ_{HqX}^{(1,3)}$. In order to study the possible excesses in the $[2,\,3]$, $[3,\, 4]$, $[4,\, 5]$ and $[5,\, 6]\GeV^2$ bins, we choose $m_\DM=1.7$, $2.0$, $2.2$ and $2.4\GeV$ as benchmarks, respectively. For the MFV couplings of these 6 operators, we choose the minimal values allowed by all the experimental constraints, as done for the operators involving two DM fields. By using these benchmark points, we also show in figure~\ref{fig:BR} the bin-averaged branching ratios of the $B^+ \to K^+ + \inv$ and $B^0 \to K^{*0} + \inv$ decays. Since the minimal values allowed by the Belle II measurement are chosen, the distributions resulting from these operators are the same for the $B^+ \to K^+ + \inv$ decay spectrum. The resulting $B^0 \to K^{*0} + \inv$ decay spectra are, however, distinguishable from each other for the operators $\mQ_{dHX}$, $\mQ_{qa}$ and $\mQ_{d\phi}$. They can also be distinguished from that resulting from the operators $\mQ_{qX}$ and $\mQ_{HqX}^{(1,3)}$, which have the same distribution. As the two-body decay kinematics is quite different from that of the three-body decay, a dedicated statistical analysis can provide more insightful information, for which we refer to ref.~\cite{Altmannshofer:2023hkn} for a recent study.

\begin{figure}[t]
  \centering
  \includegraphics[width=0.49\linewidth]{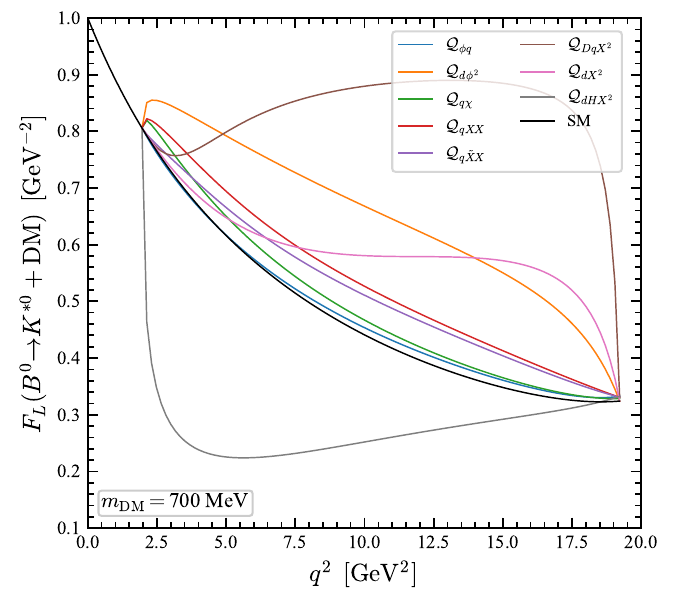}
  \includegraphics[width=0.49\linewidth]{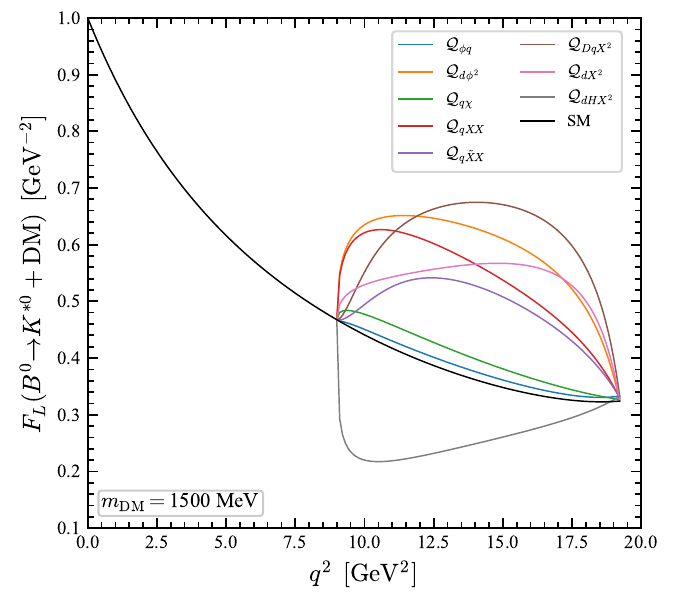}
  \caption{Longitudinal polarization fraction of the $B^0 \to K^{*0} + \inv$ decay both within the SM and in the DSMEFT with MFV hypothesis. The benchmark points used in this figure are the same as the ones in figure~\ref{fig:BR}.}
  \label{fig:FL}
\end{figure}

For the $B \to K^* +\inv$ decays, the polarization of the $K^*$ meson can be extracted via an angular analysis of its decay products. This virtue allows us to define an additional observable, the longitudinal polarization fraction $F_L \equiv ({\rm d}\Gamma_L/ {\rm d} q^2)/ ({\rm d}\Gamma/ {\rm d} q^2)$, where ${\rm d}\Gamma_L$ denotes the differetial decay width with a longitudinally polarized $K^*$. By using the same benchmark points used to investigate the branching ratios in figure~\ref{fig:BR}, our predictions for the longitudinal polarization fraction of the $B^0 \to K^{*0} + \inv$ decay are shown in figure~\ref{fig:FL}. Comparing figures~\ref{fig:BR} and \ref{fig:FL}, we can see that the longitudinal polarization fraction is complementary to the branching ratio and can, therefore, serve as an additional observable to distinguish the various DSMEFT operators. For example, in the case of $m_\DM = 1500\MeV$, the operators $\mQ_{q\chi}$ and $\mQ_{dX^2}$, while predicting quite similar $B^0 \to K^* + \inv$ decay spectra, result in quite different $F_L$. However, in the case of $m_\DM = 700 \MeV$, the operators $\mQ_{qXX}$ and $\mQ_{q \tilde{X} X}$, which are the remaining indistinguishable operators by combine the $B^+ \to K^+ + \inv$ and $B^0 \to K^* + \inv$ decay spectra, show at most around $5\%$ difference in $F_L$.

From figures~\ref{fig:S:MFV}-\ref{fig:V:MFV:2}, we can see that the benchmark points actually correspond to the minimal allowed parameter values required to explain the $B^+ \to K^+ + \inv$ excess. They are irrelevant to the $b \to d + \inv$ and $s \to d + \inv$ decays. Therefore, the predicted branching ratios and longitudinal polarization fractions of the $B^+ \to K^+ + \inv$ and $B^0 \to K^{*0} + \inv$ decays in figures~\ref{fig:BR} and \ref{fig:FL} can also be applied to the general DSMEFT case discussed in the last subsection.

\section{Conclusion}
\label{sec:conclusion}

The recent Belle II measurement of $\mB(B^+ \to K^+ \nu \bar\nu)$ is about $2.7\sigma$ higher than the SM prediction. In this work, we have deciphered the data with two different NP scenarios: the underlying quark-level $b \to s \nu \bar\nu$ transition is, besides the SM contribution, further affected by some heavy new mediators that are much heavier than the electroweak scale, or amended by an additional decay channel with light new final states that are sufficiently long-lived to escape the detector and appear as missing-energy signals. To make our analyses model-independent as much as possible, we have studied these two scenarios in the SMEFT and the DSMEFT framework, respectively. Furthermore, the flavour structures of the resulting effective operators are taken to be either of the most generic form or satisfy the MFV hypothesis, both for the quark and lepton sectors.

In the first scenario, we have focused on the implications of the MFV hypothesis. It is found that, once the MFV hypothesis is assumed for the quark sector while without any assumption on the lepton sector, only one SM-like low-energy effective operator $\mO_L^{\nu_i \nu_j}$ can be induced by the SMEFT dim-6 operators. As a consequence, the ratio of the branching fractions of any two $b \to s \nu \bar\nu$ decays remains a constant, which can be used to test the MFV hypothesis. In particular, the resulting ratio $\mB(B^+ \to K^+ \nu \bar\nu)_\MFV/\mB(B^0 \to K^{*0} \nu \bar\nu)_\MFV=0.46 \pm 0.07$ is inconsistent with the current $90\%$ CL lower limit $0.70$ derived by combining the recent Belle II measurement of $\mB(B^+ \to K^+ \nu \bar\nu)$ and the Belle upper bound on $\mB(B^0 \to K^{*0} \nu\bar\nu)$. Thus, the Belle II excess cannot be explained in the SMEFT with MFV hypothesis for the quark sector. Such a conclusion is true irrespective of the flavour structure of the lepton sector. As the uncertainty of the Belle II measurement of $\mB(B^+ \to K^+ \nu \bar\nu)$ is still quite large, we have also used all the relevant processes but without the $B^+ \to K^+ \nu \bar\nu$ decay to constrain the parameter space of the MFV parameters, and then predicted that $\mB(B^+ \to K^+ \nu \bar\nu)_\MFV < 10.5 \times 10^{-6}$ and $3.5 < \mB(B^+ \to K^+ \nu \bar\nu)_\MFV < 5.0 \times 10^{-6}$, which correspond to the cases without and with the leptonic MFV, respectively. These predictions can be further tested at the Belle II with more statistics.

In the second scenario, it is found that the Belle II excess can be accommodated by 22 DSMEFT operators with dim $\leq 6$, which include 4 operators involving scalar, 2 operators involving fermionic, 14 operators involving vector DM, as well as 2 operators involving ALP. For each operator, the current experimental constraints on the corresponding Wilson coefficient and DM (ALP) mass are derived. For most of these operators, large parts of the parameter spaces required to account for the $B^+ \to K^+ + \inv$ excess are excluded by the $B^0 \to K^{*0} + \inv$ and $B_s \to \inv$ decays. Nevertheless, none of the operators is fully excluded by the current data. The future Belle II ($5\ab^{-1}$), CEPC and FCC-ee (Tera-$Z$ phase) data on $B_s \to \phi + \inv$ and $B_s \to \inv$ decays are, however, expected to cover almost all the parameter spaces needed for explaining the Belle II excess. Furthermore, the $B^0 \to K^0 +\inv$ decay can provide an independent probe of the NP mechanism proposed for the Belle II excess. Once the MFV hypothesis is assumed, 8 out of the 22 viable DSMEFT operators do not induce the down-type FCNC interactions, and hence cannot account for the Belle II excess. For the remaining 14 operators, their MFV couplings make the $b \to d$ and $s \to d$ processes also relevant. For $\mQ_{d\phi}$, $\mQ_{qa}$, $\mQ_{qX}$ and $\mQ_{HqX}^{(1,3)}$, which all involve one DM field, the $K^+ \to \pi^+ + \inv$ decay provides the strongest constraint in the low-mass region $m_\DM \lesssim m_K-m_\pi$. For $\mQ_{q \chi}$, $\mQ_{dHX}$, $\mQ_{qX}$, $\mQ_{HqX}^{(1,3)}$ and $\mQ_{dHX^2}$, especially for the ones (e.g., $\mQ_{dHX^2}$) that make no contribution to the $B_s \to \inv$ decay, the $B^+ \to \pi^+ + \inv$ decay puts the most stringent bound in the high-mass region. In addition, the $\Upsilon(1S) \to \inv$ decay provides the strongest constraint in the high-mass region for the operator $\mQ_{\phi q}$ with $m_{\phi_2} - m_{\phi_1} < 200\MeV$. Within the parameter space allowed by all the current experimental data, the differential branching ratios (as well as the longitudinal polarization fractions) of the $B \to K^{(*)} + \inv$ decays are then studied for each viable operator. We find that the resulting prediction of the operator $\mathcal{Q}_{q\chi}=\big(\bar{q}_{p} \gamma_\mu q_{r} \big) \big(\bar{\chi} \gamma^\mu \chi \big)$ with a fermionic dark matter mass $m_\chi \approx 700\MeV$ can closely match the Belle II event distribution in the bins $2\leq q^2 \leq 7 \GeV^2$. Our numerical results also show that the longitudinal polarization fractions are complementary to the branching ratios. By combining them, all the DSMEFT operators are distinguishable from each other, except the operators $\mQ_{qXX}$ and $\mQ_{q \tilde{X} X}$ in the case of $m_X=700\MeV$.

At the Belle II with $5\ab^{-1}$ dataset, the branching ratio $\mB(B^+ \to K^+ + \inv)$ is expected to be measured with an uncertainty of $19\%$. If the central value of the current Belle II measurement remains unchanged but the uncertainty is reduced accordingly in the future, the discrepancy from the SM prediction will increase to $4\sigma$ level. At the same time, in the future, the $B^0 \to K^0 + \inv$, $B^0 \to K^{*0}+\inv$, $B^+ \to \pi^+ + \inv$, $B^+ \to \rho^+ + \inv$ and $B_s \to \inv$ decays at Belle II~\cite{Belle-II:2018jsg,Belle-II:2022cgf}, the $B_s \to \phi + \inv$ at CEPC~\cite{Li:2022tov} and FCC-ee~\cite{Amhis:2023mpj} as well as the $K \to \pi + \inv$ at NA62~\cite{NA62:2017rwk} and KOTO~\cite{Yamanaka:2012yma}, are all expected to provide complementary information about the (D)SMEFT operators as well as the validity of the MFV hypothesis studied here.

The new singlets introduced in the DSMEFT could be the DM candidates in the universe. In this case, the parameter space derived in our work could be further constrained by cosmological observations, such as the DM relic density and the Big Bang nucleosynthesis, as well as the DM direct detection~\cite{Nollett:2013pwa,Cheung:2012gi,Balazs:2014rsa,Liem:2016xpm,GAMBIT:2021rlp}. For the operators involving one new particle, their couplings should also suffer from additional constraints to ensure that the particle is sufficiently long-lived to escape the detector. In addition, the DM FCNC couplings can affect the neutral-meson mixings, such as $B_{s} - \bar B_{s}$ mixing~\cite{MartinCamalich:2020dfe,Bauer:2021mvw}. At the LHC, the large couplings of the new singlets to the third generation in MFV can lead to the DM production associated with top and bottom quarks~\cite{Lin:2013sca}. However, a complete and dedicated study of all these constraints should go beyond the EFT framework and specific UV completions are needed. Therefore, we leave these explorations for our future work.

\acknowledgments

This work is supported by the National Natural Science Foundation of China under Grant Nos.~12135006, 12075097 and 11805077, as well as by the Fundamental Research Funds for the Central Universities under Grant No.~CCNU22LJ004. XY is also supported in part by the Startup Research Funding from CCNU.

\appendix

\section{Hadronic matrix elements}
\label{eq:hadronic matrix element}

For the $B$-meson and kaon decays discussed in sections~\ref{sec:theory:nunu} and \ref{sec:theory:DM}, as well as the invisible decays of charmoniums and bottomoniums, we need the hadronic matrix elements between the vacuum and the initial hadronic states or between the initial and the final hadronic state, which can be parameterized by the decay constants or by the transition form factors. For convenience, they are recapitulated in appendices~\ref{eq:decay constant} and \ref{sec:form factor}, respectively. The matrix elements of other bilinear quark currents not mentioned explicitly are zero due to parity invariance of strong interaction.

\subsection{Decay constants}
\label{eq:decay constant}

For the $B_q$ meson ($q=u, d$), the non-vanishing annihilation matrix elements are defined as
\begin{align}
  \langle 0|\bar{q} \gamma_{\mu} \gamma_5 b | \bar{B}_q (p)\rangle = i f_{B_q} \, p_\mu,
  \quad\quad
  \langle 0|\bar{q} \gamma_{5}  b | \bar{B}_q (p) \rangle = -i \frac{m_{B_q}^2 }{\overline{m}_q(\mu)+\overline{m}_b(\mu)}f_{B_q},
\end{align}
with $\overline{m}_b(\mu)$ and $\overline{m}_q(\mu)$ denoting the $b$- and $q$-quark running masses at the scale $\mu$ defined in the $\overline{\text{MS}}$ scheme. Here $f_{B_q}$ denote the $B_q$-meson decay constants, and their up-to-date Lattice QCD values can be found in ref.~\cite{FlavourLatticeAveragingGroupFLAG:2021npn} and references therein. For the vector quarkonium $V$ (such as $\Upsilon$ and $J/\psi$), the relevant hadronic matrix elements are given by~\cite{Aloni:2017eny,Hatton:2021dvg}
\begin{align}
  \langle 0 | \bar{q} \gamma^\mu q | V(p,\varepsilon) \rangle &= f_V m_V \varepsilon^\mu,\nonumber \\[0.4em]
  \langle 0 | \bar{q} \sigma^{\mu\nu} q | V(p,\varepsilon) \rangle &= if_V^T(\mu) \big ( \varepsilon^\mu p^\nu - \varepsilon^\nu p^\mu\big),
\end{align}
where $\varepsilon^{\mu}$ denotes the polarization vector of the vector quarkonium, and $f_V^{(T)}$ is the decay constant. The matrix element of the axial-tensor current can be related to that of the tensor current, while the ones of all the other currents are zero. In the numerical analysis, we use the values of $f_\Upsilon$ in ref.~\cite{Hatton:2021dvg}, $f_{J/\psi}$ in ref.~\cite{Hatton:2020qhk}, and $f_V^T(2\GeV)/f_V$ in ref.~\cite{Konig:2015qat}.

\subsection{Transition form factors}
\label{sec:form factor}

The transition form factors for a $B$ meson decaying into a pseudoscalar meson $P$ (e.g., $B\to K$) are defined by~\cite{Beneke:2000wa,Ball:2004ye}
\begin{align}
  c_P \langle P (p^\prime)|\bar{q} \gamma^{\mu} b | \bar{B} (p) \rangle &= f_+(q^2) 
  \left(P^\mu - \frac{m_B^2-m_P^2}{q^2} q^\mu \right) + f_0(q^2) \frac{m_B^2-m_P^2}{q^2}q^\mu, \nonumber\\[0.4em]
  c_P \langle P (p^\prime)|\bar{q} b | \bar{B} (p) \rangle &= \frac{m_B^2-m_P^2}{\overline{m}_b(\mu)-\overline{m}_q(\mu)} f_0(q^2), \nonumber\\[0.4em]
  c_P \langle P (p^\prime)|\bar{q} \sigma^{\mu\nu} b (p) | \bar{B} (p) \rangle &= i \frac{P^\mu q^\nu - P^\nu q^\mu}{m_B+m_P} f_T(q^2),
\end{align}
where $q=p-p^\prime$, $P=p+p^\prime$, and $c_P=-\sqrt{2}$ for $\pi^0$ and $c_P=1$ otherwise. For a $B$ meson decaying into a vector meson $V$, the relevant transition form factors are introduced by the following Lorentz decompositions of the bilinear quark current matrix elements~\cite{Beneke:2000wa,Ball:2004rg}:
\begin{align}\label{eq:FF:tensor}
  c_V \langle V(p^\prime,\varepsilon^*)|\bar{q} \gamma_{\mu} b | \bar{B} (p) \rangle &= -\epsilon_{\mu\nu\rho\sigma} \varepsilon^{*\nu} P^\rho q^\sigma \frac{V(q^2)}{m_B+m_V}, \nonumber\\[0.4em]
  c_V \langle V(p^\prime,\varepsilon^*)|\bar{q} \gamma^{\mu} \gamma^5 b | \bar{B} (p) \rangle &= i\varepsilon^{*\mu}  (m_B+m_V) A_1(q^2) -iP^\mu \frac{\varepsilon^{*}\cdot q}{m_B+m_V}A_2(q^2) \nonumber\\
  &\qquad -i q^\mu(\varepsilon^*\cdot q)\frac{2 m_V}{q^2}\left[ A_3(q^2)-A_0(q^2) \right],
  \nonumber\\[0.4em]
  c_V \langle V(p^\prime,\varepsilon^*)|\bar{q} \gamma^5 b | \bar{B} (p) \rangle &= -2i\frac{m_V}{\overline{m}_b(\mu)+\overline{m}_q(\mu)} (\varepsilon^* \cdot q) A_0(q^2),
  \nonumber\\[0.4em]
  c_V \langle V(p^\prime,\varepsilon^*)|\bar{q} \sigma_{\mu\nu} q^\nu b | \bar{B} (p) \rangle 
  &= -i\epsilon_{\mu\nu\rho\sigma} \varepsilon^{*\nu} P^\rho q^\sigma T_1 (q^2),
  \nonumber\\[0.4em]
  c_V \langle V(p^\prime,\varepsilon^*)|\bar{q} \sigma_{\mu\nu} q^\nu \gamma^5 b | \bar{B} (p) \rangle 
  &= \left[ (m_B^2-m_V^2) \varepsilon^*_\mu - (\varepsilon^*\cdot q) P_\mu \right] T_2(q^2) \nonumber\\
  &\qquad +  (\varepsilon^*\cdot q) \left (q_\mu - \frac{q^2}{m_B^2-m_V^2} P_\mu\right ) T_3(q^2),
\end{align}
where $\varepsilon$ denotes the polarization vector of the vector meson, and $c_V=-\sqrt{2}$ for $\rho^0$ and $c_V=1$ otherwise. We adopt the convention $\epsilon^{0123}=+1$ and $\sigma_{\mu\nu}=i/2\left[\gamma_\mu,\gamma_\nu\right]$. With the above parameterizations, the absence of dynamical singularities at $q^2 = 0$ implies that $f_+(0) = f_0(0)$ and $A_0(0) = A_3(0)$. In addition, the algebraic relation between $\sigma_{\mu\nu}$ and $\sigma_{\mu\nu}\gamma_5$, $\sigma_{\mu\nu}\gamma_5=i/2\epsilon_{\mu\nu\alpha\beta}\sigma^{\alpha\beta}$, gives rise to the identity $T_1(0) = T_2(0)$. It is also noted that the form factor $A_3$ is redundant and can be written as a linear combination of $A_1$ and $A_2$~\cite{Ball:2004rg}
\begin{align}
A_3(q^2)=\frac{(m_B+m_V)}{2 m_V} A_1(q^2) - \frac{m_B-m_V}{2 m_V}A_2(q^2).
\end{align}
The hadronic matrix elements of the tensor and axial-tensor currents can also be decomposed as~\cite{Khodjamirian:2020btr}
\begin{align}
    c_V\langle V(p^\prime,\varepsilon^*)|\bar{q} \sigma^{\mu\nu} b | \bar{B}(p) \rangle
    =& i\epsilon_{\mu\nu\rho\sigma} \left[ P^\rho\varepsilon^{*\sigma} - \frac{m_B^2 - m_V^2}{q^2} q^\rho \varepsilon^{*\sigma} + \frac{\varepsilon^* \cdot q}{q^2} q^\rho P^\sigma \right] T_1(q^2) \nonumber\\
    & + i\epsilon_{\mu\nu\rho\sigma} \left[\frac{m_B^2 - m_V^2}{q^2} q^\rho \varepsilon^{*\sigma} + \frac{\varepsilon^* \cdot q}{q^2} P^\rho q^{\sigma} \right] T_2(q^2) \nonumber\\
    & + i\epsilon_{\mu\nu\rho\sigma} \frac{\varepsilon^* \cdot q}{m_B^2 - m_V^2} P^\rho q^{\sigma} T_3(q^2) ,
    \nonumber\\[0.2cm]
    c_V\langle V(p^\prime,\varepsilon^*)|\bar{q} \sigma^{\mu\nu} \gamma^5 b | \bar{B}(p) \rangle
    =& \bigg[(\varepsilon^{*\mu} P^\nu - P^\mu \varepsilon^{*\nu}) 
    + \frac{m_B^2 - m_V^2}{q^2} (q^\mu \varepsilon^{*\nu} - \varepsilon^{*\mu} q^\nu) \nonumber\\
    & + \frac{\varepsilon^*\cdot q}{q^2} (P^\mu q^{\nu} - q^{\mu} P^\nu)\bigg]T_1(q^2) \nonumber\\
    &\hspace{-1.3cm} + \left[\frac{m_B^2 - m_V^2}{q^2} (\varepsilon^{*\mu} q^\nu - q^\mu \varepsilon^{*\nu}) + \frac{\varepsilon^*\cdot q}{q^2} (q^{\mu} P^\nu - P^\mu q^{\nu})\right]T_2(q^2) \nonumber\\
    & + \frac{\varepsilon^*\cdot q}{m_B^2 - m_V^2} \left(q^{\mu} P^\nu - P^\mu q^{\nu}\right) T_3(q^2),
\end{align}
from which the forms given in eq.~\eqref{eq:FF:tensor} can be obtained by multiplying both sides by $q_{\nu}$. All the $B_s \to P\,(V)$ and $K \to \pi$ form factors can be defined similarly.

Numerically, we use the values obtained in ref.~\cite{Ball:2004ye} for the $B \to P$ and in ref.~\cite{Ball:2004rg} for the $B \to V$ transition form factors. As the Lattice QCD results are still not available for all of these form factors, especially for the tensor ones (see ref.~\cite{FlavourLatticeAveragingGroupFLAG:2021npn} and references therein), we use instead the calculations based on the light-cone sum rule approach~\cite{Ball:2004ye,Ball:2004rg}. For the $K\to \pi$ vector and scalar form factors, the results obtained in refs.~\cite{Bernard:2006gy,Bernard:2009zm} are adopted, while the value in ref.~\cite{Baum:2011rm} is used for the $K \to \pi$ tensor form factor.

For the operator $\mQ_{DdX^2}$ defined in eq.~(\ref{eq:DSMEFT:V:2}), the relevant form factors are not available in the literature so far. However, when the Wilson coefficient $\mC_{DdX^2}$ is assumed to be hermitian, they can be derived from the above form factors. Taking the $\bar B \to \bar K^0$ transition as an example, we have the relation $\langle \bar K^0(p^\prime) \lvert (\partial^\mu \bar s) \gamma^\nu P_L b + \bar s \gamma^\nu P_L (\partial^\mu b) \rvert \bar B(p) \rangle = - i q^\mu \langle \bar K^0(p^\prime) \lvert  \bar s \gamma^\nu P_L b \rvert \bar B(p) \rangle $.

\bibliographystyle{JHEP}

\begingroup
\setlength{\bibsep}{5pt}
\bibliography{ref}
\endgroup

\end{document}